\documentclass[aps,prd,twocolumn,showpacs,nofootinbib,floatfix,superscriptaddress,showpacs]{revtex4-1}
\makeatletter

\newcommand{\Rmnum}[1]{\expandafter\@slowromancap\romannumeral #1@}
\makeatother
\usepackage[colorlinks,linkcolor=blue,anchorcolor=blue,citecolor=blue,urlcolor=blue,breaklinks=true]{hyperref}

\usepackage{epsfig}
\usepackage{natbib}
\usepackage{epstopdf}
\usepackage{mathrsfs}
\usepackage{slashed}
\usepackage{amsmath}
\usepackage{verbatim}
\usepackage{graphicx}
\usepackage{amssymb}
\usepackage{psfrag}
\usepackage{array}
\usepackage{lipsum}

\newcommand{\ud}{\mathrm{d}}

\begin{document}

\baselineskip=10pt

\author{Yi-Lun Du}
\email{yldu@fias.uni-frankfurt.de}
\affiliation{Frankfurt Institute for Advanced Studies, Giersch Science Center, D-60438 Frankfurt am Main, Germany}
\affiliation{Institut f$\ddot{u}$r Theoretische Physik, Goethe Universit$\ddot{a}$t Frankfurt, D-60438 Frankfurt am Main, Germany}
\affiliation{Department of Physics, Nanjing University, Nanjing 210093, China}
\affiliation{Department of Physics and Technology, University of Bergen, 5007 Bergen, Norway}

\author{Kai Zhou}
\affiliation{Frankfurt Institute for Advanced Studies, Giersch Science Center, D-60438 Frankfurt am Main, Germany}

\author{Jan Steinheimer}
\affiliation{Frankfurt Institute for Advanced Studies, Giersch Science Center, D-60438 Frankfurt am Main, Germany}

\author{Long-Gang Pang}
\affiliation{Nuclear Science Division, Lawrence Berkeley National Laboratory, Berkeley, California 94720, USA}
\affiliation{Physics Department, University of California, Berkeley, CA 94720, USA}
\affiliation{Key Laboratory of Quark and Lepton Physics (MOE) and Institute of Particle Physics, Central China Normal University, Wuhan 430079, China}

\author{Anton Motornenko}
\affiliation{Frankfurt Institute for Advanced Studies, Giersch Science Center, D-60438 Frankfurt am Main, Germany}
\affiliation{Institut f$\ddot{u}$r Theoretische Physik, Goethe Universit$\ddot{a}$t Frankfurt, D-60438 Frankfurt am Main, Germany}

\author{Hong-Shi Zong}
\affiliation{Department of Physics, Nanjing University, Nanjing 210093, China}
\affiliation{Nanjing Proton Source Research and Design Center, Nanjing 210093, China}
\affiliation{Department of physics, Anhui Normal University, Wuhu, Anhui 241000, China}

\author{Xin-Nian Wang}
\affiliation{Nuclear Science Division, Lawrence Berkeley National Laboratory, Berkeley, California 94720, USA}
\affiliation{Physics Department, University of California, Berkeley, CA 94720, USA}
\affiliation{Key Laboratory of Quark and Lepton Physics (MOE) and Institute of Particle Physics, Central China Normal University, Wuhan 430079, China}

\author{Horst St\"{o}cker}
\affiliation{Frankfurt Institute for Advanced Studies, Giersch Science Center, D-60438 Frankfurt am Main, Germany}
\affiliation{Institut f$\ddot{u}$r Theoretische Physik, Goethe Universit$\ddot{a}$t Frankfurt, D-60438 Frankfurt am Main, Germany}
\affiliation{GSI Helmholtzzentrum f$\ddot{u}$r Schwerionenforschung, D-64291 Darmstadt, Germany}

\title{Identifying the nature of the QCD transition in relativistic collision of heavy nuclei with deep learning}


\begin{abstract}

Using deep convolutional neural network (CNN), the nature of the QCD transition can be identified from the final-state pion spectra from hybrid model simulations of heavy-ion collisions that combines a viscous hydrodynamic model with a hadronic cascade ``after-burner". Two different types of equations of state (EoS) of the medium are used in the hydrodynamic evolution. The resulting spectra in transverse momentum and azimuthal angle are used as the input data to train the neural network to distinguish different EoS. Different scenarios for the input data are studied and compared in a systematic way. A clear hierarchy is observed in the prediction accuracy when using the event-by-event, cascade-coarse-grained and event-fine-averaged spectra as input for the network, which are about 80\%, 90\% and 99\%, respectively. A comparison with the prediction performance by deep neural network (DNN) with only the normalized pion transverse momentum spectra is also made. High-level features of pion spectra captured by a carefully-trained neural network were found to be able to distinguish the nature of the QCD transition even in a simulation scenario which is close to the experiments.

\end{abstract}


\maketitle
\section{Introduction}\label{sec:intro}
The dynamics of the strong interactions between quarks and gluons, governing the properties of hot and dense nuclear matter, can be described by the theory of QCD. It predicts that, if the temperature of strongly-interacting matter becomes large enough, a new state of matter is formed in which quarks and gluons can roam freely and are not confined in the hadrons anymore. This state of matter is called the quark-gluon plasma (QGP). 
Lattice QCD has established that the transition from a hadron gas to the QGP is a smooth crossover at a high temperature $T\thicksim 140-180$ MeV and low net baryon density~\cite{Aoki:2006we,Aoki:2006br,Karsch:2001cy}. A variety of theoretical models, such as the Dyson-Schwinger equations model~\cite{qin2011phase,fischer2013propagators,JHEP.1407.014,shi2016continuum}, the (Polyakov loop-) Nambu-Jona-Lasinio model~\cite{costa2008thermodynamics,costa2009qcd,PhysRevD.77.114028,fu20082+,PhysRevD.88.114019} and the quark-meson coupling model \cite{schaefer2007phase,nickel2009inhomogeneous,skokov2011quark} also predict the existence of a first-order phase transition that occurs at low temperature and moderate to large net baryon densities. 

Relativistic heavy ion experiments have been carried out at the SIS18~\cite{TAHIR200516}, at the AGS~\cite{rai1999results} and at the SPS~\cite{lansberg2012quarkonium} in the fixed target mode and at the Relativistic Heavy Ion Collider (RHIC) \cite{adams2005experimental} as well as at the Large Hadron Collider (LHC)~\cite{muller2012first} in the collider mode. The forthcoming Facility for Anti-proton and Ion Research (FAIR)~\cite{friman2011cbm,ablyazimov2017challenges} and the Nuclotron-based Ion Collider fAcility (NICA)~\cite{sissakian2009nuclotron} will provide unprecedented intensities and luminosities for future studies. The main goal of these large experiments is to search for signals for the QCD phase transition and study the properties of QGP in nucleus-nucleus collisions.
Due to the transience of the heavy ion collision dynamics, the QCD medium bulk properties can't be directly observed in experiment. A strategy to identify the signals of QGP is to compare sophisticated model simulations with varying parameter sets and different equations of state (with and without a phase transition) with experimental data such as particle spectra and correlation functions. Currently some observables, for example, anisotropic flow~\cite{luzum2008conformal,Hofmann:1976dy,stoecker1986high,Hung:1994eq}, directed flow \cite{Rischke:1995pe,Brachmann:1999xt} and fluctuations of particle multiplicities \cite{Stephanov:1998dy,Stephanov:2008qz,Koch:2008ia,Steinheimer:2012gc}, are conjectured as most sensitive to the appearance of a phase transition. However, no disentangled mapping between these observables and this specific bulk property of the QCD medium from others, has been obtained so far. Then it's necessary to call for modern data analysis methods like Bayesian analysis or the deep neural network approach.

The Bayesian analysis~\cite{pratt2015constraining,bernhard2016applying,bernhard2019bayesian} applies a global fitting to a set of different observables for parameter estimation. In Ref.~\cite{bernhard2019bayesian}, the crossover type EoS was employed in the hybrid hydrodynamic framework and the event-averaged experimental data (e.g. particle yields, momentum distribution and flow) were used to infer the temperature-dependent shear and bulk viscosity of nuclear matter and other parameters at the same time. These estimated temperature-dependent viscosities are their marginal distributions by integrating out other parameters, respectively. One way to constrain these bulk properties of nuclear matter better is to fit more data or make use of more information from data. On one hand, one can employ higher-dimensional raw data instead of the integrated one to fit. On the other hand, the event-by-event fluctuation may contain more information as well.

In this work, we will explore the feasibility of identifying QCD EoS from event-by-event high-dimensional raw hadron spectra in high energy nucleus-nucleus collisions using the tools and techniques of Deep Learning (DL). DL was developed to capture highly-correlated features from big data~\cite{schmidhuber2015deep,lecun2015deep}. It has achieved tremendous success in a wide variety of applications, like image processing, natural language processing, computer vision, medical imaging, medical information processing, and other interesting fields. These have inspired physicists to adopt the technique to tackle physical problems of great complexity. A lot of progresses have been made in nuclear physics~\cite{pang2018equation,Huang:2018fzn,chien2018probing,PhysRevC.94.024907, utama2016nuclear,haake2017machine,PhysRevLett.114.202301}, lattice field theory~\cite{zhou2019regressive,Urban:2018tqv,Mori:2017nwj,Shanahan:2018vcv,Tanaka:2017niz}, particle physics~\cite{baldi2014searching,baldi2015enhanced,searcy2016determination,barnard2017parton,moult2016new}, astrophysics~\cite{schawinski2017generative,erdmann2018deep,PhysRevD.98.023019} and condensed matter physics~\cite{mehta2014exact,carrasquilla2017machine,carleo2017solving,torlai2016learning,broecker2016machine,ch2017machine,you2018machine}. 

For our exploration with DL method here, the purpose is to find out a disentangled mapping between observed final raw spectra and the EoS type for the medium. We vary different parameters, including shear viscosity, equilibration time, freeze-out temperature, etc., to enforce the neural network to explore if it can find a direct mapping from event-by-event high-dimensional raw spectra to the EoS type which can be immune to other parameters' `interference' in certain ranges. As long as we can find such a mapping, it’s straightforword to infer information about the EoS type from the measured data in experiment as the detector simulation or calibration is also considered for further study.

The great advantage of the DL method over conventional ones is its ability to extract hidden features from highly dynamical, rapidly evolving and complex non-linear systems, like in relativistic heavy ion collisions. Conventional observables rely on human's design and are usually low-dimensional projections of the high-dimensional raw data. When one uses only part of these projected information to constrain the properties of nuclear matter, the estimated value are prone to be dependent on the specific model setup (e.g. other untuned parameters in the fitting) and the chosen observables. Instead DL methods can be used to explore distinct mappings and to construct observables from the full high-dimensional raw data for the classification task at hand. Recently, a deep CNN classifier was developed as an effective ``EoS-meter", an excellent tool for revealing the nature of the QCD transition with a high predictive accuracy $\thicksim95\%$ in hadron spectra from a pure hydrodynamic study~\cite{pang2018equation}. 

The present work studies the performance of a CNN to identify the EoS trained and tested with hadron spectra from a more realistic simulation of heavy ion collisions. The generalizability of the method is explored by considering well established dynamics in the state-of-the-art simulation models. First of all, the hadronic rescattering, after the hydrodynamics evolution, is taken into account in the simulation via a hadronic cascade. Consequently, the event-by-event final-state pion spectra are discrete instead of smooth as in hydrodynamic simulations. Secondly, the resonance decays are included, which also contribute to the pion spectra. Due to the finite number of particles, the discrete event-by-event pion spectra will have significant fluctuations that might overwhelm correlations one is looking for. We will develop modified DL-tools with CNN to identify the EoS in this more complex and more realistic dynamic scenario. 

This paper is organized as follows: Sec.~\ref{sec:model} introduces the hybrid simulation model. Sec.~\ref{sec:neural network} discusses the neural network and the methods of the data pre-processing. Sec.~\ref{sec: train-test} presents the performance of the trained CNN in different scenarios and comparisons with that of a fully-connected deep neural network (DNN). Finally, Sec.~\ref{sec: summary} summarizes the results and gives the conclusions. ~\ref{appendix:neural network} gives the details of the neural network structure.~\ref{appendix: data} shows the simulated data and predictive performance on testing datasets by the trained neural network. ~\ref{appendix: traditional observables} visualizes the training datasets in~\ref{appendix: data} with traditional observables.

\section{Micro-Macro hybrid model of relativistic heavy-ion collisions}\label{sec:model}

The modeling of relativistic heavy-ion collision is mostly done by following a ``standard prescription'' for the spatio-temporal evolution of the collision dynamics. The initial state of the matter right after the violent collision is described by the ``color glass condensate", which consists of frozen primordial gluons and is assumed to isotropize within 1 fm/c~\cite{florkowski2013anisotropic,kovchegov2012quantum,Stoecker:2015zea,Stocker:2015nka,Vovchenko:2016ijt}. These gluons may evolve rapidly in accordance with the classical Yang-Mills equation. A few fm/c later, they can achieve approximate local thermal equilibrium~\cite{Xu:2014ega,Zhou:2017zql} and may exist briefly as a Yang-Mills gluon plasma, which may quickly expand nearly isentropically due to the high initial temperature. The total entropy and energy are not yet distributed over quark-anti-quark degrees of freedom. Subsequently, quarks are produced by gluon-gluon collisions~\cite{kovchegov2012quantum,Stoecker:2015zea,Stocker:2015nka,Vovchenko:2016ijt}, forming a strongly coupled quark-gluon plasma (sQGP). The dynamical evolution of that QGP can be described approximately by macroscopic dissipative hydrodynamics~\cite{heinz2010early,romatschke2010new,song2008causal,schenke2011elliptic,pang2012effects,pang2015analytical}. Viscous corrections are included to describe some of the remaining deviation from local isotropy and thermal equilibrium. The EoS of the hot QGP medium, the constitutive element used to close the hydrodynamic equations, is one crucial input. As the medium expands and cools quasi-isentropically, the quark-gluon fluid will go through a smooth crossover, or hypothetically in this work as a control experiment, a first order phase transition. The nature of the QCD transition strongly affects the hydrodynamic evolution~\cite{stoecker2005collective}. Different forms of transitions are associated with different pressure gradients which consequently lead to different expansion rates. As the matter becomes more dilute, it will form an expanding non-equilibrium hadronic matter with important final state effects. For instance, final absorption of the products of the resonance decays in the hadronic matter can substantially change the yields of the hadrons observed by the experimental detectors. This evolution of the hadronic matter can be successfully described by microscopic hadron cascade models~\cite{bass1998microscopic,bleicher1999relativistic,Steinheimer:2017vju}.

To generate the data for the training of the CNN, we use the iEBE-VISHNU hybrid model~\cite{shen2016iebe}, which can perform event-by-event simulations of relativistic heavy-ion collisions at different energies. Major components of this hybrid model include an initial condition generator (SuperMC), a (2+1)D second-order event-by-event viscous hydrodynamic simulator (VISHNew), a particle sampler (iSS) and a hadron cascade ``afterburner" simulator (UrQMD). 

This hybrid model uses either the Monte-Carlo Glauber (MC-G)~\cite{broniowski2009glissando,alver2008phobos,loizides2015improved} or the Monte-Carlo Kharzeev-Levin-Nardi (MCKLN) ~\cite{kharzeev2005onset,kharzeev2005color} model to generate the fluctuating initial conditions in the SuperMC module. The collision centrality can be set up as needed, based on the assumption that, on average, the final charged hadron multiplicity, ${\ud}N_{\mathrm{ch}}/{\ud}y$, is directly proportional to the initially produced total entropy in the transverse plane ${\ud}S/{\ud}y|_{y=0}$. The effect of viscous heating will cause a spread in the final ${\ud}N_{\mathrm{ch}}/{\ud}y$, which is considered small (2-3\%) for a given ${\ud}S/{\ud}y|_{y=0}$. 

\begin{figure}[b]
\centering
\includegraphics[width=0.5\textwidth]{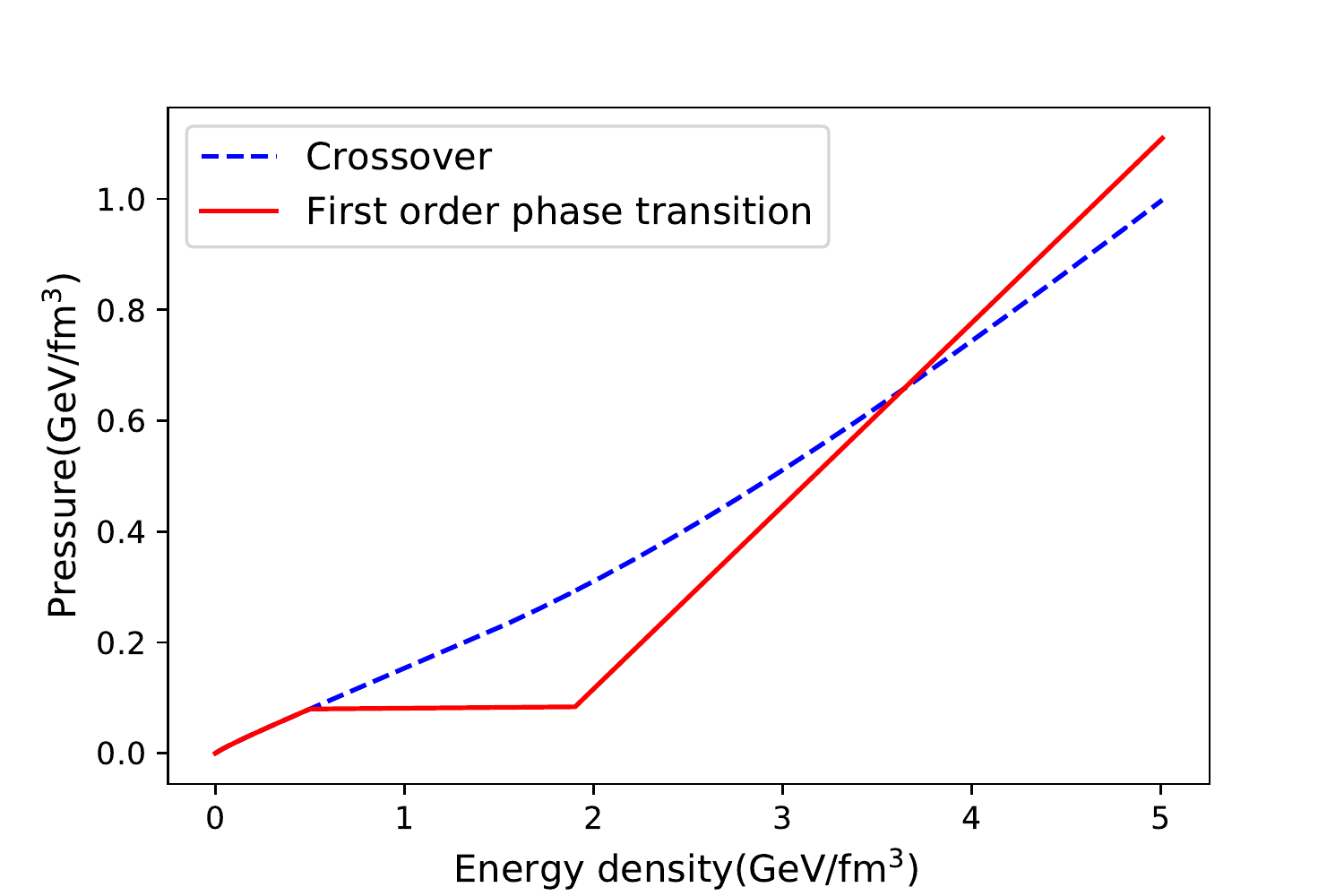}
\caption{Two different EoSs are implemented in the hydrodynamic simulation, as functions of the energy density. A crossover, based on a lattice QCD parametrization is compared with a first order phase transition with a transition temperature $T_c=165$ MeV, obtained by a Maxwell construction. It is assumed that the baryon-chemical potential is exactly $\mu_B=0$ throughout the whole simulation.}
\label{eos}
\end{figure}

The simulation with the hydrodynamic package VISHNew uses two different EoSs: (1) the crossover type EoS, based on a lattice-QCD parametrization~\cite{huovinen2010qcd}, denoted as L-EOS; (2) the first order type EoS with a Maxwell construction~\cite{sollfrank1997hydrodynamical} between a hadron resonance gas and an ideal gas of quarks and gluons, as Q-EOS. The transition temperature is $T_c=165$ MeV. These two EoSs are depicted in Fig.~\ref{eos}. 

After the hydrodynamic evolution, the fluid fields are projected via the Cooper-Frye formula into particles, which will then be further propagated in a hadronic cascade, the Ultrarelativistic Quantum Molecular Dynamics (UrQMD) model. 
In UrQMD, a non-equilibrium transport model, resonance decays and hadronic rescatterings are included in the simulation. In contrast to the hydrodynamic evolution, which is governed by the conservation of energy and momentum with the EoS, shear viscosity $\eta$, bulk viscosity $\xi$, particles are assumed to be in asymptotic states and the trajectories are given by straight-lines between the collisions in the hadronic cascade. The hadronic cascade evolution is not deterministic since the processes involve certain randomness, e.g., scattering angle, scattering probability and decay probabilities. Furthermore, the effects of finite number of particles, i.e., thermal fluctuations, are included since the cascade propagates the discrete particles instead of the average densities.

This hybrid model with some adjustable parameters can fit experimental data on final hadron spectra. These parameters include: the equilibration time $\tau_0$, which defines the point when the local thermal equilibration is reached and the hydrodynamics evolution starts, the ratio of the shear viscosity to the entropy density $\eta/s$, and the freeze-out temperature $T_{sw}$, which defines the switch from the hydrodynamic evolution to the hadronic cascade.

We vary the model parameters in the generation of the training datasets to allow the neural network to capture the intrinsic features encoded in the EoS, instead of those biased by the specific setup of other physical uncertainties. This would require many events simulations for hundreds of different parameter combinations and centrality selections, to make sure that the neural network gains a sufficient generalizabilty. However, in practice this is impossible. Hence we focus on systematic changes of these parameters and study the performance of the network whence it reaches the boundary of these parameter values.

\section{Neural network and data pre-processing}\label{sec:neural network}
In Ref.~\cite{pang2018equation}, the DL-tool engine with a CNN has been shown to classify successfully the EoS in pure hydrodynamical simulations, on an event-by-event basis with a $\thicksim 95\%$ accuracy. To apply this strategy to real experimental data, it's crucial to perform realistic simulations with hadronic ``after-burner" and resonance decays. In the present paper, the DL-tool engine is constructed for more realistic simulations of heavy ion collisions. The CNN architecture used here is similar to that discussed in Ref.~\cite{pang2018equation}. We refer to that paper for technical details. An introduction to this new CNN network is presented in detail in Fig.~\ref{cnn} in~\ref{appendix:neural network}.

The input $\rho(p_T,\Phi) \equiv dN_{\pi}/dy dp_T d\Phi$ to this neural network is a histogram of the number of pions with 24 $p_T$-bins and 24 $\Phi$-bins. $p_T$ denotes the transverse momenta of observed pions in the final state, while $\Phi$ denotes the azimuthal angles. Only pions with $p_T\leq 2 \mathrm{GeV}$, rapidity $|y|\leq 1$ and $\Phi\in [0, 2\pi]$ are accepted and accounted in the histogram.

In general, training or learning algorithms benefit a lot from pre-processing of the datasets. The input to the neural network used here, pion spectra $\rho(p_T,\Phi)$, is a $24\times24$ matrix. One refers to each matrix element as one ``feature" and each matrix as one ``sample". The pre-processing of the input data can be applied in a feature-wise (per feature) or sample-wise (per input sample) manner.

In the feature-wise standardization, the input $\rho(p_T,\Phi)$ of all the training samples are pre-processed in a sample-interdependent manner. Each feature is subtracted with the mean over all training samples and is divided by their standard deviation. In this way, all features are centered around zero and have variances of the same order. Thus it is prevented that one feature with larger variance dominates the objective function over other features. The transformation is saved and then will be applied in the testing samples. With this standardization, the testing data should be simulated in one of the same collision systems as the training data, since the multiplicity in different collision systems differ a lot.

In the sample-wise standardization, or min-max normalization, the inputs $\rho(p_T,\Phi)$ are pre-processed in a sample-independent manner. Each $24\times24$ matrix can be rescaled to have a zero mean and a unit variance, or to a specific range, such as $[-\frac{1}{2}, \frac{1}{2}]$, respectively. The latter choice is used in Ref.~\cite{pang2018equation} with success. 

Our training results show that feature-wise standardization does always perform better than the other two sample-wise methods. Hence we will show in the following only the results of the feature-wise standardization. 

\section{Training and testing results}\label{sec: train-test}
A systematic analysis of the performance of the above described CNN is presented for hybrid modeling for relativistic heavy-ion collisions. Here an important aspect is the generalizability of the trained CNN model in the testing stage. The overfitting of the network to the training data will be checked on the validation data which are generated with the same physical parameter set in modeling the training data. The testing is performed on the testing datasets which are generated with different physical parameter sets in modeling the training data. The generalizability of the CNN model with respect to different physical parameter sets is studied systematically. In the previous study with pure hydrodynamics~\cite{pang2018equation}, the training data are generated with a viscous (3+1)D hydrodynamics model, CLVisc~\cite{pang2012effects}, with AMPT initial conditions~\cite{Lin:2004en}, while the testing data are generated with a viscous (2+1)D hydrodynamics model, VISHNew, with Monte-Carlo Glauber initial conditions, which are used in a hybrid model in this work for the training data generation instead. However, here we find that, even in the pure hydrodynamic study, reversing the simulation models for training and testing data generation will obtain a testing accuracy only about 70\%, from which we suspect some
superiority of (3+1)D hydrodynamics model with AMPT initial conditions over other ones. Thus in this work, we would not be able to discuss the generalizability of the CNN model with respect to different hybrid simulation models.

\subsection{Hybrid model with late transition to cascade}
\label{late transition}
The CNN in the previous study~\cite{pang2018equation} was directly trained using primordial pion spectra, obtained from a numerical integration of the Cooper-Frye formula over the freeze-out hypersurface in the hydrodynamics. In such a scenario, one neglects the fluctuations due to the finite number of hadrons. In addition, a significant portion of pions originating from resonance decays also need to be taken into account. In this section, we study the influence of the aforementioned effects on the predictive power of the CNN.
To see the influence of the finite number of particles and resonance decays, we first assume a late transition from hydrodynamics to the UrQMD cascade by taking the switching temperature the same value as the hydrodynamics freeze-out temperature used in Ref.~\cite{pang2018equation}, $T_{sw}=137$ MeV. In this scenario, the duration and influence of the hadronic cascade are significantly diminished and we are left with the effects of the finite number of particles and resonance decays as compared to the pure hydrodynamics modeling. 

\subsubsection{Event-by-event input, switch at $T_{sw}=137$ MeV}
In this sub-scenario, the event-by-event pion spectra $\rho(p_T, \phi)$ are taken as the input to the CNN. 12 training datasets are generated by the iEBE-VISHNU hybrid model with the fluctuating MC-Glauber initial condition and 6 different fine centrality bins with 1\% width in the centrality range 0-60\% in two collision systems, respectively. We set the ratio of the shear viscosity to entropy density as $\eta/s=0.08$ and $0.00$, the equilibration time as $\tau_0=0.5$ and $0.4~\mathrm{fm/c}$ in the collision systems Pb+Pb $\sqrt{s_{NN}}=2.76~\mathrm{TeV}$ and Au+Au $\sqrt{s_{NN}}=200~\mathrm{GeV}$, respectively. The details of the datasets are shown in Tabs.~\ref{data 2760} and~\ref{data 200} in~\ref{appendix: data}. About 44000 events with two different EoSs are generated in total. Fig.~\ref{fig-PTV2_spectra_ebe} in~\ref{appendix: traditional observables} shows the event-by-event normalized $p_T$ spectra and the elliptic flow $v_2$ as a function of $p_T$ of these training datasets with two EoSs. These two one-dimensional traditional observables are non-distinguishable by the human eye with respect to the EoSs. Thus it's not trivial to identify the EoS from just final-state pion $p_T$ spectra. The negative elliptic flow $v_2$ in Fig.~\ref{fig-PTV2_spectra_ebe} shows that there are great fluctuations in the event-by-event spectra.

The validation accuracy is found to be about 83.5\% after 1000 epochs training. This validation accuracy indicates that high-level correlations are extracted from the two-dimensional pion spectra $\rho(p_T, \phi)$ to identify the EoS. However, it is significantly lower than that in pure hydrodynamics modeling~\cite{pang2018equation}, where a validation accuracy up to 99\% was obtained. This implies that the fluctuations due to the finite number of particles and resonance decays overwhelm some correlation information from the early dynamics to the final-state particle spectra and thus result in the ``overlap" between these two types of event-by-event spectra with different EoSs, which hinders the discrimination between them. 
 
\subsubsection{Cascade-coarse-grained input, switch at $T_{sw}=137$ MeV}
To mitigate the effect of fluctuation due to the finite number of particles and resonance decays, we average the pion spectra over a certain number of events. In the model simulations one can repeat the hadronic cascade for any number of times for the same hydrodynamic evolution. Then the pion spectra averaged over these simulations are taken as the input for training, which will be called ``cascade-coarse-grained input". We would like to find out whether such an event averaging will improve the network performance due to the statistics enhancement or worsen it due to the information loss. 

In this sub-scenario, 2 training datasets are generated by the iEBE-VISHNU hybrid model with the fluctuating MC-Glauber initial condition in the centrality range 0-50\%. The details are shown in Tab.~\ref{training137cascade} in~\ref{appendix: data}. In total, 15747 events are generated with two different EoSs. The hadronic cascade is repeated 30 times after each hydrodynamics evolution. The spectra averaged over these 30 events are taken as the input to the network. 
The validation accuracy with these cascade-coarse-grained spectra $\rho_c(p_T, \phi)$ can achieve about $92\%$. One can see that such averaging over cascade-stage is beneficial in identifying the EoS information in early dynamics from the final-state particle spectra. This means that the statistics matters a lot for using particle spectra to decode the EoS information.

\subsubsection{Event-fine-averaged input, switch at $T_{sw}=137$ MeV}
One drawback of the above average procedure is that the separation of collision dynamics into hydrodynamic and hadronic cascade stage is purely theoretical. Thus from a realistic point of view, an averaging procedure based on experimentally controllable event filtering is preferable. In this sub-scenario, spectra are averaged within the same fine centrality bin (with 1\% width) instead, which will be called ``event-fine-averaged input" in the following. To be specific, we average the spectra of 30 random events within the same fine centrality bin in Tabs.~\ref{data 2760} and~\ref{data 200} as the input to the network to accumulate the statistics. Fig.~\ref{fig-PTV2_spectra_fine} in~\ref{appendix: traditional observables} shows the 30-events-fine-averaged normalized $p_T$ spectra and the elliptic flow $v_2$ as a function of $p_T$ of these training datasets with two EoSs. These two one-dimensional traditional observables are still not distinguishable by eye. By comparing with the corresponding event-by-event observables as shown in Fig.~\ref{fig-PTV2_spectra_ebe}, one can see that the fluctuations are significantly reduced in the 30-events-fine-averaged spectra. This manner of averaging reduces the fluctuations from the initial conditions besides that from hadronic cascade and resonance decays. Consequently, a surprisingly obvious improvement for the CNN performance in classifying the two types of EoS is made. The validation accuracy reaches about $99\%$ with the 30-events-fine-averaged spectra $\rho_a(p_T, \phi)$ after 1000 epochs training, a value similar to that in the pure hydrodynamic case~\cite{pang2018equation}. In principal, one can include more datasets generated in different fine centrality bins for training. However, we confirm that it's enough to use the datasets simulated only in 6 representative fine centrality bins as in Tabs.~\ref{data 2760} and~\ref{data 200}, respectively, for training, since the predictive performances on the datasets simulated in other unselected fine centrality bins are as high as the training accuracy. This demonstrates that non-trivial high-level correlations which are independent of the centrality bins are learned by the neural network.

After the training with validation, the trained network is confronted with the testing data, which are generated with different physical parameter sets in simulations to explore the network's generalizability. In Tabs.~\ref{test137Glb} and~\ref{test137KLN} in~\ref{appendix: data} we show the predictive performance of the neural network trained with the 30-events-fine-averaged spectra.
A testing accuracy 95\% on average is obtained on the testing data simulated in the centrality range 0-50\% with MC-Glauber or MCKLN initial conditions. This evidently demonstrates that the trained neural network is robust against different model setups such as initial conditions, $\tau_0$, $\eta/s$ and $T_{sw}$ in a range between [130, 142] MeV. We observe a slight centrality dependence of the predictive accuracy in the collision system Pb+Pb  $\sqrt{s_{NN}}=2.76~\mathrm{TeV}$, which decreases for more peripheral events.

\subsubsection{A hierarchy of the accuracy in the above sub-scenarios}
\begin{figure}[tb]
\centering
\includegraphics[width=0.50\textwidth]{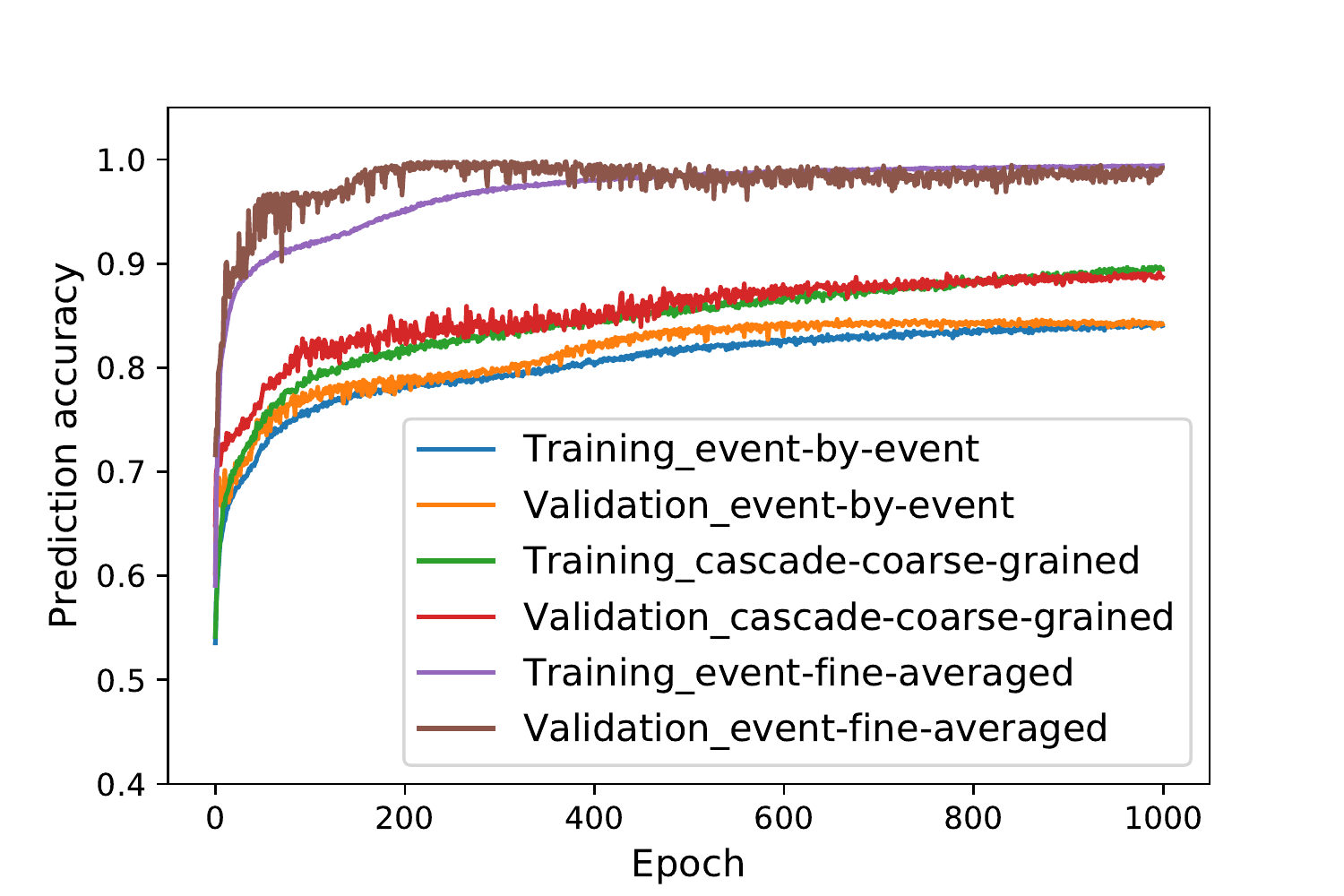} 
\includegraphics[width=0.50\textwidth]{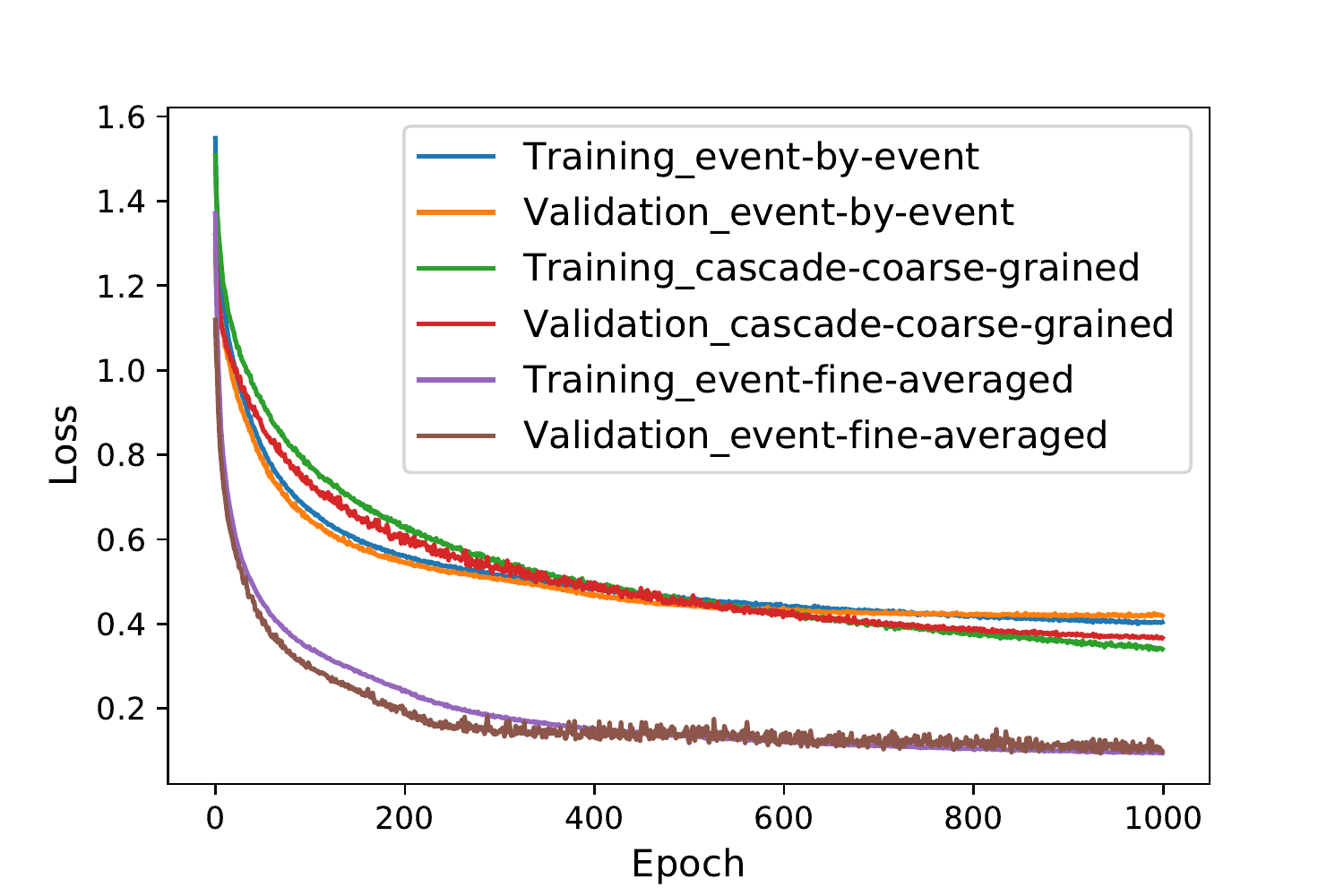} 
\caption{Training and validation accuracy (upper panel) and loss (lower panel) in three different sub-scenarios with switching temperature $T_{sw}=137$ MeV. These three sub-scenarios refer to the 30-events-fine-averaged spectra (purple and brown), the cascade-coarse-grained spectra (red and green) as well as the event-by-event spectra (blue and orange).}
\label{fig-accuracyloss}
\end{figure}
Fig.~\ref{fig-accuracyloss} shows the training and validation accuracy (upper panel) and loss (lower panel), respectively, by the CNN with the same setup for the first 1000 epochs in three aforementioned sub-scenarios. In each sub-scenario, training and validation accuracy (loss) are still close after 1000 epochs training, which implies that over-fitting is avoided. Besides, the network has not been sufficiently trained in the cascade-coarse-grained sub-scenario after 1000 epochs as the accuracy (loss) is still increasing (decreasing). 

A clear hierarchy of the prediction accuracy is observed when the averaging is performed over more and more stages of the simulated dynamics. The CNN with event-by-event spectra gives the lowest accuracy, while the one with the 30-events-fine-averaged spectra gives the highest one, which is as high as in the pure hydrodynamic study~\cite{pang2018equation}.

\subsection{Hybrid model with early transition to cascade}
\label{early transition}
The scenario with early transition from hydrodynamics to hadronic cascade in hybrid modeling is in accordance with a widely used choice of the switching temperature $T_{sw}>150$ MeV. This scenario is different from the one discussed in the previous subsection in two aspects. Firstly, the higher switching temperature decreases the contribution from the primordial pions which are directly emitted from the hydrodynamic evolution, and increases the contribution from resonance decays. Secondly, the elongated duration of the hadronic cascade stage may further blur out the imprint of the phase transition encoded in the final-state particle spectra. In the following, we will study how a higher switching temperature affects the performance of the CNN in three aforementioned sub-scenarios, respectively.

\subsubsection{Event-by-event input, switch at $T_{sw}>150$ MeV}
In this sub-scenario, 9 training datasets are generated  by the iEBE-VISHNU hybrid model with the fluctuating MC-Glauber initial condition in the centrality range 0-50\%. The switching temperature is $T_{sw} = 160$ MeV. Two different values for the equilibration time $\tau_0$ and the ratio of shear viscosity to entropy $\eta/s$ are used in the simulations. The details are shown in Tabs.~\ref{data-2760-160} and~\ref{data-200-160} in~\ref{appendix: data}. In total, about 60000 events are generated with two different EoS types. 

The validation accuracy is found to be about $78\%$ for the CNN trained with these event-by-event spectra as input. This validation accuracy is lower than that in the sub-scenario with late transition (switching temperature $T_{sw} = 137$ MeV). This decrease in the validation accuracy can be understood as a result of the increased contribution from resonance decays and the elongated duration of the hadronic rescattering.

\subsubsection{Cascade-coarse-grained input, switch at $T_{sw}>150$ MeV}
In this sub-scenario, the cascade-coarse-grained pion spectra $\rho_c(p_T, \phi)$ are taken as the input to the CNN. 2 training datasets are generated in analogy to the previous late transition case, by the iEBE-VISHNU hybrid model with the fluctuating MC-Glauber initial condition in the centrality range 0-50$\%$ with the hadronic cascade simulated 30 times individually after each hydrodynamic evolution. The switching temperature $T_{sw}$ is set to be 155 or 160 MeV. The details are shown in Tab.~\ref{training155cascade} in~\ref{appendix: data}. About 24000 events with two different EoSs are generated in total. The validation accuracy is found to be 87.5$\%$ at most, which is also lower than that in previous sub-scenario with late transition to cascade.

4 testing datasets are generated in this sub-scenario as shown in Tab.~\ref{test_sample} in~\ref{appendix: data} in the centrality range 0-50\%. Both MC-Glauber and MCKLN initial conditions are used, and simulation parameters are varied from the training datasets to check the generalizability of the CNN. After training and validating the neural network, the testing accuracy on these datasets is $83\%$ on average, which is slightly lower than the validation accuracy.

\subsubsection{Event-fine-averaged input, switch at $T_{sw}>150$ MeV}
In this sub-scenario, the 30-events-fine-averaged spectra for training is explored with the switching temperature $T_{sw} = 160$ MeV. This input is generated by the average over the spectra of 30 independent events within the same fine centrality bins (with 1\% width) shown in Tabs.~\ref{data-2760-160} and~\ref{data-200-160}. The validation accuracy can also reach up to $99\%$ in this sub-scenario as in the previous late transition one. The testing accuracy is up to $95\%$ on average on the testing datasets as shown in Tab.~\ref{test-data-central-155} in the~\ref{appendix: data}. We also observe a slight centrality dependence of the predictive accuracy in the collision system Au+Au  $\sqrt{s_{NN}}=200~\mathrm{GeV}$, which decreases for more peripheral events. 

It’s also interesting to further check the performance of the neural network on the testing datasets which employ temperature-dependent shear viscosities. Here taking this sub-scenario for example, we evaluated the network's prediction accuracy on the testing datasets in Tab.~\ref{test-data-central-155-viscosity} where four temperature-dependent shear viscosities are employed in hybrid simulations as shown in Fig.~\ref{Fig: Temperature_dependent_viscosity} (labelled as 1-4, respectively). The first two are taken from Ref.~\cite{niemi2011influence}, while the last two are taken from the Bayesian analysis estimations~\cite{bernhard2019bayesian,bernhard2016applying}, respectively). The results show that the performance is robust against the setup of these temperature-dependent shear viscosities as compared with Tab.~\ref{test-data-central-155}.

\subsection{Comparison with fully-connected deep neural network}
As already discussed in subsection~\ref{late transition}, the event-by-event and 30-events-fine-averaged normalized $p_T$ spectra and elliptic flow $v_2$ with two different EOS from all centrality bins in Tabs.~\ref{data 2760} and~\ref{data 200}, as shown in Figs.~\ref{fig-PTV2_spectra_ebe} and~\ref{fig-PTV2_spectra_fine}, respectively, are non-distinguishable within the range of event-by-event fluctuations. However, one can observe that the peaks of the normalized $p_T$ spectra with Q-EOS are higher than that with L-EOS on the whole. In Figs.~\ref{fig-PTV2_spectra_ebe_onebin},~\ref{fig-PTV2_spectra_fine_onebin} and~\ref{fig-PTV2_spectra_mean_onebin} in~\ref{appendix: traditional observables}, we show the event-by-event, 30-events-fine-averaged and all-events-fine-averaged normalized $p_T$ spectra (left panel) and elliptic flow $v_2$ (right panel) solely from centrality bin 14-15\% in Pb+Pb collision $\sqrt{s_{NN}}=2.76$ TeV in Tab.~\ref{data 2760}, respectively. Within the same centrality bin one can see that the all-events-fine-averaged normalized $p_T$ spectra are distinguishable with respect to different EOSs, 30-events-fine-averaged normalized $p_T$ spectra are almost distinguishable from certain $p_T$ bins, while the event-by-event normalized $p_T$ spectra are still not. In Fig.~\ref{fig-PTV2_spectra_mean} in~\ref{appendix: traditional observables}, we show the all-events-fine-averaged normalized $p_T$ spectra (upper left panel) and elliptic flow $v_2$ (upper right panel) as well as the first (lower left panel), second (lower middle panel) and third (lower right panel) derivatives of the normalized $p_T$ spectra from all centrality bins in Tabs.~\ref{data 2760} and~\ref{data 200}. One can see that these all-events-fine-averaged normalized $p_T$ spectra are not distinguishable again by the human eye. Their derivatives are also helpful to distinguish the EoS in certain $p_T$ bins, which might lead us to construct novel observables from normalized $p_T$ spectra in the future. Inspired with this observation, we use the normalized $p_T$ spectra as the input to a fully-connected DNN to distinguish the EOSs as a first try. In this case, the normalized $p_T$ spectra are regarded as a whole instead of isolated points at each $p_T$ bin as regarded by the human eye, and high-level correlations including but not limited to high-order derivatives can be extracted supervisely.

We train a fully-connected DNN~\footnote{This fully-connected DNN consists of two hidden dense layers of size 128 and 256, respectively, and each is followed by a dropout~\cite{srivastava2014dropout} (with a rate of 0.5) and PReLu activation layer~\cite{he2015delving}. These two dense layers are initialized with ``He normal" initializer~\cite{he2015delving} and constrained with L2 regularization~\cite{ng2004feature}.} with the event-by-event normalized $p_T$ spectra from all centrality bins in Tabs.~\ref{data-2760-160} and~\ref{data-200-160} as the input. The validation accuracy is about 74\%, which is below that by CNN with two-dimensional spectra, about 78\%. Here the correlations are not very strong in both cases due to the fluctuations from the particlization and "afterburner". When the 30-events-fine-averaged normalized $p_T$ spectra are taken as the input instead, the validation accuracy is about 97\%, which is also a little below that by CNN with two-dimensional spectra, about 99\%. Here the correlations are very strong in both cases. As for the testing accuracy, CNN with two-dimensional spectra outperforms fully-connected DNN with one-dimensional spectra by about $8\%$ with 30-events-fine-averaged spectra. Apparently, in the above cases, fully-connected DNN with one-dimensional normalized $p_T$ spectra can capture the main correlations, while CNN with two-dimensional spectra performs better and improves the generalizability. 

When the event-by-event normalized $p_T$ spectra from all centrality bins in Tabs.~\ref{data 2760} and~\ref{data 200} with $T_{sw}=137$ MeV and in Tabs.~\ref{data-2760-160} and~\ref{data-200-160} with $T_{sw}=160$ MeV are taken as the input to the fully-connected DNN, the validation accuracy is about $62\%$, which is much lower than that by CNN with two-dimensional spectra, about $69\%$. This shows that when physical parameters in the simulation model vary a lot in the generation of the training data, the normalized $p_T$ spectra are more difficult to distinguish and CNN with two-dimensional spectra will outperform fully-connected DNN with one-dimensional normalized $p_T$ spectra.

\section{Summary and Conclusion}
\label{sec: summary}
\begin{figure}[bt]
\centering
\includegraphics[width=0.50\textwidth]{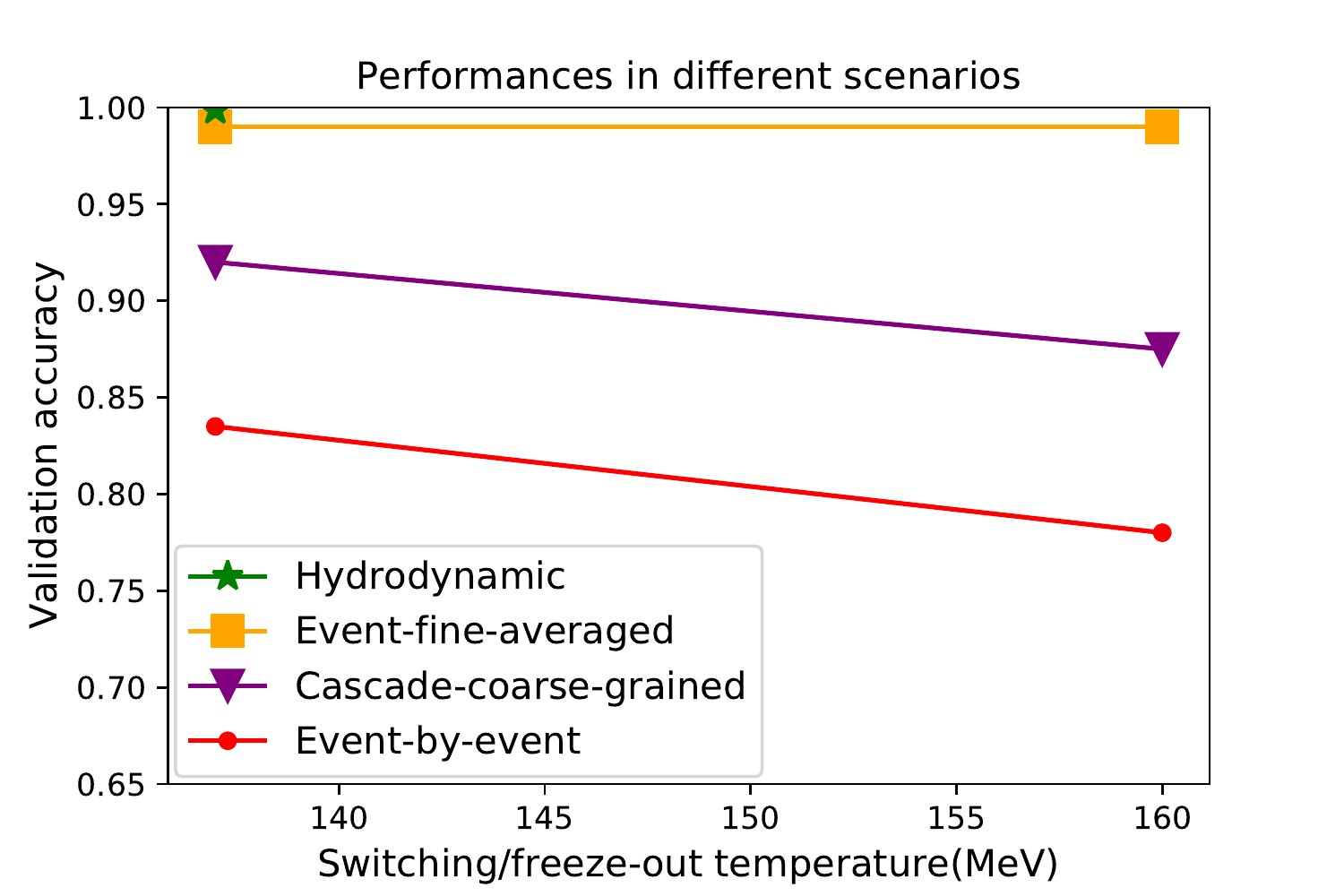} 
\caption{Comparison between the validation accuracy in all the different sub-scenarios studied. The green star depicts the pure hydrodynamic result~\cite{pang2018equation}. The orange square, the purple triangle and the red filled circle symbols depict the results for the 30-events-fine-averaged, cascade-coarse-grained and event-by-event spectra, respectively, in different switching temperatures.}
\label{Fig: Performance}
\end{figure}
We extended a previous exploratory study on identifying EoS in the modeling of heavy ion collisions from hadron spectra using DL technique~\cite{pang2018equation}. In this extended study, we consider more realistic hybrid modeling for heavy-ion collisions, where hadronic cascade ``afterburner" with finite number of particles and resonance decays are properly taken into account. In the hybrid modeling the final-state particle spectra are histograms containing large fluctuations and thus are different from those in the previous study~\cite{pang2018equation}, which are smooth hadron spectra from Cooper-Frye prescription with perfect statistics. Fig.~\ref{Fig: Performance} summarizes the predictive performances on the validation datasets in the above exploratory studies of different sub-scenarios.

We have demonstrated that, after the hydrodynamic evolution, stochastic particlization, hadronic cascade and resonance decays, the information about EoS in early dynamics is preserved in the final-state pion spectra, from the perspectives of deep CNN, as shown in Fig.~\ref{Fig: Performance}. The event-by-event input for the network can reveal the EoS-type information with about 80\% classification accuracy in binary classification setup. 

The downward trend for the performance of network in validation with respect to the switching temperature in Fig.~\ref{Fig: Performance}, implies that more stochasticity from the resonance decays and the elongated hadronic cascade will diminish the correlation between the EoS information in the early dynamics and the final-state particle spectra. This is in accordance with the common physical interpretation.

Finally, the hierarchy of the validation accuracy in different sub-scenarios in Fig.~\ref{Fig: Performance} shows that proper enhancement of statistics and reduction of fluctuations from either the final hadronic dynamics or together with the initial conditions in the input data are found to facilitate the revealing of the EoS information by the network from final-state particle spectra.

In conclusion, deep CNN can decode the imprint of the EoS in hydrodynamic evolution (encoded within the phase transition dynamics) on the final-state pion spectra from heavy-ion collisions. The good performance of the network does demonstrate that this ``EoS-encoder" works. The fingerprint of the early dynamics of the bulk matter is not washed out by the evolution even when stochasticity is increased due to the hadronization and sequential hadron dynamics. Deep CNN provides an effective decoding method to extract high-level correlations from two-dimensional final-state pion spectra, which are immune to different physical factors, such as centrality bins. In relatively simple cases, fully-connected deep neural network can also identify the EoS from normalized pion $p_T$ spectra with close validation accuracy as CNN does, which can lead us to discover new observables sensitive to EoS from normalized pion $p_T$ spectra. The generalizability of the learned features with respect to other simulation models also depends on the simulation model for the training data generation. In the present study, the training data is generated with well tested iEBE-VISHNU (VISHNew+UrQMD) hybrid model. In the future we will explore how to capture the features which can be generalized to the testing data from other models as well as experimental data. Possible applications of the framework developed here can be extended to classifying fluctuating initial conditions, extracting transport coefficients of QCD matter, analysis of real experimental data filtering and pre-processing, and detector calibration.

\section*{Acknowledgement}
Y.D. thanks Chun Shen for the helpful illustrations of the usage of iEBE-VISHNU package and Volodymyr Vovchenko for helpful discussions. This work is supported by the Helmholtz Graduate School HIRe for FAIR (Y. D. and A. M.) , by the F\&E Programme of GSI Helmholtz Zentrum f$\ddot{\mathrm{u}}$r Schwerionenforschung GmbH, Darmstadt (Y. D.), by the Giersch Science Center (Y. D.), by the Walter Greiner Gesellschaft zur F$\ddot{\mathrm{o}}$rderung der physikalischen Grundlagenforschung e.V., Frankfurt (Y. D.), by the AI grant of SAMSON AG, Frankfurt (Y. D., K. Z. and J. S.), by the BMBF under the ErUM-Data project (K. Z. and J. S.), by the NVIDIA Corporation with the donation of NVIDIA TITAN Xp GPU for the research (K. Z. and J. S.), and by the Judah M. Eisenberg Laureatus Chair by Goethe University and the Walter Greiner Gesellschaft, Frankfurt (H.St.), by Trond Mohn Foundation under Grant No. BFS2018REK01 (Y. D.), by National Natural Science Foundation of China under Grant Nos.11475085, 11535005, 11690030 (Y. D. and H. Z.) and 11221504 (X.-N.W.), and National Major state Basic Research and Development of China under Grant Nos. 2016Y-FE0129300 (Y. D. and H. Z.) and 2014CB845404 (X.-N.W.), and the U.S. Department of Energy under Contract Nos. DE-AC02-05CH11231 (L. P. and X.-N.W.), and the U.S. National Science Foundation (NSF) under Grant No. ACI-1550228 (JETSCAPE) (L. P. and X.-N.W.).

\appendix
\section{Neural network structure}
\label{appendix:neural network}
\begin{figure*}[htb]
\centering
\includegraphics[width=0.98\textwidth]{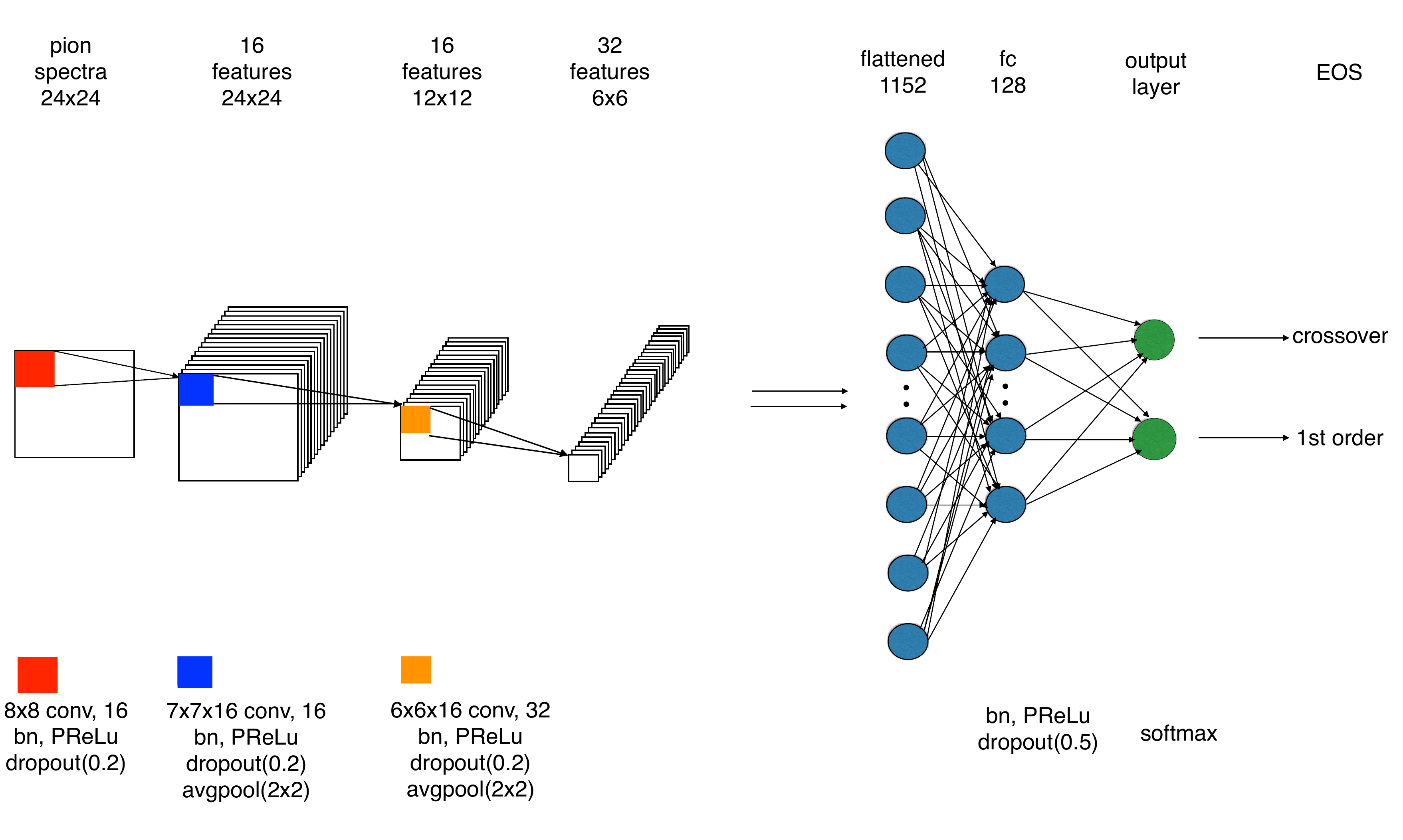} 
\caption{The architecture of our convolution neural network (CNN) for identifying the QCD transition type by using pion spectra with 24 transverse momentum $p_T$ bins and 24 azimuthal angle $\Phi$ bins.}
\label{cnn}
\end{figure*}
Fig.~\ref{cnn} shows the neural network architecture. We use three convolutional layers and one subsequent fully-connected layer. All the convolutional layers and the fully-connected one are followed by a batch normalization~\cite{ioffe2015batch}, PReLu activation~\cite{he2015delving}, dropout~\cite{srivastava2014dropout} (with a rate of 0.2 and 0.5, respectively) and average pooling (of pool size $2\times2$, following last two convolutional layers only) layer, one by one. There are 16, 16, 32 filters of size $8\times8$, $7\times7$ and $6\times6$, respectively, in these three convolutional layers, scanning through the input $\rho(p_T, \Phi)$, or the previous layers, and creating 16, 16, 32 features of size $24\times24$, $24\times24$, $12\times12$, respectively. The weight and bias matrix of these convolutional layers are initialized with ``He normal" initializer~\cite{he2015delving}, i.e. truncated normal distribution with zero mean and standard deviation $\sqrt{2 /N_\mathrm{in}}$ where $N_\mathrm{in}$ is the number of input units in the weight tensor. They are constrained with L2 regularization~\cite{ng2004feature}. Each neuron in a convolutional layer does connect only locally to a small chunk of neurons in the previous layer by a convolution operation. This is a key reason for the success of the CNN architecture. Dropout, batch normalization, PReLU and L2 regularization, all work together to prevent overfitting, which will generate model-parameter-dependent features from the training dataset and thus hinder the generalizability of the method. The resulting 32 features of size $6\times6$ from the last average pooling layer are flattened and connected to a 128-neuron fully-connected layer. The output layer is another fully-connected layer with softmax activation and 2 special neurons which indicate the type of the EoS. There are overall 203194 trainable and 120 non-trainable parameters in the present neural network.

The supervised learning is performed in tackling this binary classification task with the L-EOS case, labeled by (1, 0), and the Q-EOS case, labeled by (0, 1). The difference between the true label and the predicted label from the two output neurons is quantified by the cross entropy~\cite{kullback1951information}, which plays the role of the loss function $l(\theta)$, where $\theta$ are the trainable parameters of the neural network. The training minimizes the loss function by updating $\theta\to\theta-\delta\theta$. Here $\delta\theta=\alpha\partial l(\theta)/\partial\theta$, where $\alpha$ is the learning rate, with initial value 0.0001, which is adaptively changed by the AdaMax method~\cite{kingma2014adam}.

The architecture is built by Keras with a Theano backend. The training datasets are fed into the network in batches with an empirically selected size of 128. One traversal of all the batches in the training datasets is called one epoch. The training datasets are reshuffled before each epoch to speed-up the convergence. The neural network is trained with 1000 epochs. The model parameters are saved to a new checkpoint whenever a smaller validation loss is encountered.

\section{Collection of the training data and predictions on the testing data}
\label{appendix: data}
\begin{table}[htbp]
\centering
\begin{tabular}{|c|c|c|}
\hline
\multicolumn{3}{|c|}{TRAINING DATASETS 1} \\
\hline
Centrality bin   &L-EOS  & Q-EOS \\
\hline
 4\%-5\%       &2539   & 2540  \\
\hline
14\%-15\%    &1022  & 1024 \\
\hline
20\%-21\%    &2814  & 2816 \\
\hline
30\%-31\%    &2560  & 2560 \\
\hline
40\%-41\%    &1024  & 1024 \\
\hline
50\%-51\%    &896  & 1024 \\
\hline
\end{tabular}
\caption{Training datasets 1: numbers of event-by-event spectra $\rho(p_T, \Phi)$ computed by the iEBE-VISHNU hybrid model with the MC-Glauber initial conditions in the centrality range 0-60\%. The ratio of shear viscosity to entropy density $\eta/s=0.08$. The equilibration time $\tau_0=0.5~\mathrm{fm/c}$. The switching temperature $T_{sw}=137$ MeV. The collision system is Pb+Pb at $\sqrt{s_{NN}}=2.76~\mathrm{TeV}$.} \label{data 2760}
\end{table}

\begin{table}[htbp]
\centering
\begin{tabular}{|c|c|c|}
\hline
\multicolumn{3}{|c|}{TRAINING DATASETS 2} \\
\hline
Centrality bin &L-EOS  & Q-EOS \\
\hline
  0\%-1\%     &979   & 1024 \\
\hline
10\%-11\%   &2560   & 2560  \\
\hline
20\%-21\%   &1024  &1024  \\
\hline
30\%-31\%   &1024  &1024  \\
\hline
40\%-41\%  &2560  & 2560\\
\hline
50\%-51\%  &2816  & 2816\\
\hline
\end{tabular}
\caption{Training datasets 2: numbers of event-by-event spectra $\rho(p_T, \Phi)$ computed by the iEBE-VISHNU hybrid model with the MC-Glauber initial conditions in the centrality range 0-60\%. The ratio of shear viscosity to entropy density $\eta/s=0.00$. The equilibration time $\tau_0=0.4~\mathrm{fm/c}$. The switching temperature $T_{sw}=137$ MeV. The collision system is Au+Au at $\sqrt{s_{NN}}=200~\mathrm{GeV}$.} \label{data 200}
\end{table}

\begin{table*}[htbp]
\centering
\begin{tabular}{| >{\centering}p{2.0cm} | >{\centering}p{2.1cm} | >{\centering}p{1.4cm} | >{\centering}p{1.0cm} | >{\centering}p{1.2cm}| >{\centering}p{1.4cm} | >{\centering}p{1.2cm}|  >{\centering\arraybackslash}p{1.3cm}|}
\hline
\multicolumn{8}{|c|}{TRAINING DATASETS 3} \\
\hline
Centrality bin & $\sqrt{s_{NN}}$ [TeV] & Ini. Cond. & $\tau_0$ $(\mathrm{fm/c})$ &$\eta/s$  & $T_{sw}$ [MeV] & L-EOS  & Q-EOS  \\
\hline
0\%-50\%   &Au+Au 0.2  & MC-G   &0.4 & 0.16    & 137 &  3990   & 4096  \\
\hline
0\%-50\%   &Pb+Pb 2.76 & MC-G   &0.6 & 0.08    & 137 & 3830  & 3835
\\
\hline
\end{tabular}
\caption{Training datasets 3: numbers of cascade-coarse-grained spectra $\rho_c(p_T, \Phi)$ computed by the iEBE-VISHNU hybrid model with the MC-Glauber initial conditions in the centrality range 0-50\%.} \label{training137cascade}
\end{table*}

\begin{table*}[!htbp]
\centering
\begin{tabular}{| >{\centering}p{1.6cm} | >{\centering}p{2.1cm} | >{\centering}p{1.1cm} | >{\centering}p{0.9cm} | >{\centering}p{0.7cm}| >{\centering}p{0.9cm} | >{\centering}p{1.2cm}|>{\centering}p{1.3cm} | >{\centering\arraybackslash}p{1.4cm}|}
\hline
\multicolumn{9}{|c|}{PREDICTIVE ACCURACY FOR TESTING DATASETS 1} \\
\hline
Centrality bin & $\sqrt{s_{NN}}$ [TeV] & Ini. Cond. & $\tau_0$ $(\mathrm{fm/c})$ &$\eta/s$  & $T_{sw}$ [MeV] & L-EOS  & Q-EOS & Accuracy \\
\hline
15\%-16\%   &Au+Au 0.2  & MC-G   &0.4  & 0.00 & 141 & 512 & 512 & 89.1\% \\
\hline
15\%-16\%   &Au+Au 0.2  & MC-G   &0.4  & 0.00 & 140 & 2560 & 2560 & 95.6\% \\
\hline
45\%-46\%   &Au+Au 0.2  & MC-G   &0.6  & 0.12 & 130 &  1024 &1024 & 100\% \\
\hline
7\%-8\%     &Pb+Pb 2.76  & MC-G  &0.6  & 0.12    & 130 & 1280 & 1279 & 99.8\% \\
\hline
17\%-18\%   &Pb+Pb 2.76  & MC-G  &0.6  & 0.12    & 130 & 2560 & 2560 & 98.1\% \\
\hline
25\%-26\%   &Pb+Pb 2.76  & MC-G  &0.6  & 0.12    & 130 & 2560 & 2560 & 97.4\% \\
\hline
25\%-26\%   &Pb+Pb 2.76  & MC-G  &0.6  & 0.16    & 130 & 1024 &1024 & 97.8\% \\
\hline
\end{tabular}
\caption{Predictive accuracy on the testing datasets 1: 30-events-fine-averaged spectra $\rho_a(p_T, \Phi)$ generated with MC-Glauber initial conditions and different $\sqrt{s_{NN}}$, $\eta/s$, $\tau_0$, and $T_{sw}$ in the centrality range 0-50\%.} \label{test137Glb}
\end{table*}

\begin{table*}[!htbp]
\centering
\begin{tabular}{| >{\centering}p{1.6cm} | >{\centering}p{2.1cm} | >{\centering}p{1.3cm} | >{\centering}p{0.9cm} | >{\centering}p{0.7cm}| >{\centering}p{0.9cm} | >{\centering}p{1.2cm}|>{\centering}p{1.3cm} | >{\centering\arraybackslash}p{1.4cm}|}
\hline
\multicolumn{9}{|c|}{PREDICTIVE ACCURACY FOR TESTING DATASETS 2} \\
\hline
Centrality bin & $\sqrt{s_{NN}}$ [TeV] & Ini. Cond. & $\tau_0$ $(\mathrm{fm/c})$ &$\eta/s$  & $T_{sw}$ [MeV] & L-EOS  & Q-EOS & Accuracy \\
\hline
15\%-16\%   &Au+Au 0.2 & MCKLN  &0.6 & 0.12    & 137& 512 &256 & 98.6\% \\
\hline
35\%-36\%   &Au+Au 0.2 & MCKLN  &0.6 & 0.12    & 142& 896 &896 & 99.4\% \\
\hline
10\%-11\%   &Pb+Pb 2.76 & MCKLN  &0.6 & 0.12    & 142 & 150 &150 & 100\% \\
\hline
25\%-26\%   &Pb+Pb 2.76 & MCKLN  &0.6 & 0.12    & 137 & 256 &256 & 84.4\% \\
\hline
\end{tabular}
\caption{Predictive accuracy on the testing datasets 2: 30-events-fine-averaged spectra $\rho_a(p_T, \Phi)$ generated with MCKLN initial conditions and the different $\sqrt{s_{NN}}$, $\eta/s$, $\tau_0$, and $T_{sw}$ in the centrality range 0-40\%.} \label{test137KLN}
\end{table*}

\begin{table*}[!htbp]
\centering
\begin{tabular}{|c|c|c|}
\hline
\multicolumn{3}{|c|}{TRAINING DATASETS 4} \\
\hline
Centrality bin & L-EOS  & Q-EOS  \\
\hline
15\%-16\%   & 2560& 2560\\
\hline
20\%-21\%   & 2560 & 2560 \\
\hline
34\%-35\%   & 3840 & 3840  \\
\hline
44\%-45\%   & 3840 & 3840 \\
\hline
\end{tabular}
\caption{Training datasets 4: numbers of event-by-event spectra $\rho(p_T, \Phi)$ computed by the iEBE-VISHNU hybrid model with the MC-Glauber initial conditions in the centrality range 0-50\%. The ratio of shear viscosity to entropy density $\eta/s=0.08$. The equilibration time $\tau_0=0.6~\mathrm{fm/c}$. The switching temperature $T_{sw}=160$ MeV. The collision system is Pb+Pb at $\sqrt{s_{NN}}=2.76~\mathrm{TeV}$.}  \label{data-2760-160}
\end{table*}

\begin{table*}[!htbp]
\centering
\begin{tabular}{|c|c|c|}
\hline
\multicolumn{3}{|c|}{TRAINING DATASETS 5} \\
\hline
Centrality bin  & L-EOS  & Q-EOS  \\
\hline
10\%-11\%   & 2560  & 2560   \\
\hline
15\%-16\%   & 2560 & 2560 \\
\hline
25\%-26\%   &  2560 & 2560 \\
\hline
34\%-35\%  &  3840 & 3840   \\
\hline
44\%-45\%   &  3840 & 3840  \\
\hline
\end{tabular}
\caption{Training datasets 5: numbers of event-by-event spectra $\rho(p_T, \Phi)$ computed by the iEBE-VISHNU hybrid model with the MC-Glauber initial conditions in the centrality range 0-50\%. The ratio of shear viscosity to entropy density $\eta/s=0.16$. The equilibration time $\tau_0=0.4~\mathrm{fm/c}$. The switching temperature $T_{sw}=160$ MeV. The collision system is Au+Au at $\sqrt{s_{NN}}=200~\mathrm{GeV}$.} \label{data-200-160}
\end{table*}

\begin{table*}[!htbp]
\centering
\begin{tabular}{| >{\centering}p{2.0cm} | >{\centering}p{2.1cm} | >{\centering}p{1.4cm} | >{\centering}p{1.0cm} | >{\centering}p{1.2cm}| >{\centering}p{1.4cm} | >{\centering}p{1.2cm}|  >{\centering\arraybackslash}p{1.3cm}|}
\hline
\multicolumn{8}{|c|}{TRAINING DATASETS 6} \\
\hline
Centrality bin & $\sqrt{s_{NN}}$ [TeV] & Ini. Cond. & $\tau_0$ $(\mathrm{fm/c})$ &$\eta/s$  & $T_{sw}$ [MeV] & L-EOS  & Q-EOS  \\
\hline
0\%-50\%   &Au+Au 0.2 & MC-G   &0.4  & 0.16    & 155 &  4608 & 4608  \\
\hline
0\%-50\%   &Au+Au 0.2 & MC-G   &0.4  & 0.00    & 155 &  3072 & 3072   \\
\hline
0\%-50\%   &Au+Au 0.2  & MC-G   &0.4 & 0.16    & 160 &  9724 & 9724  \\
\hline
0\%-50\%   &Pb+Pb 2.76 & MC-G   &0.6 & 0.08    & 155 &   5770 & 5521 \\
\hline
\end{tabular}
\caption{Training datasets 6: numbers of cascade-coarse-grained spectra $\rho_c(p_T, \Phi)$ computed by the iEBE-VISHNU hybrid model with the MC-Glauber initial conditions  in the centrality range 0-50\%.} \label{training155cascade}
\end{table*}

\begin{table*}[!htbp]
\centering
\begin{tabular}{| >{\centering}p{1.6cm} | >{\centering}p{2.1cm} | >{\centering}p{1.3cm} | >{\centering}p{0.9cm} | >{\centering}p{0.7cm}| >{\centering}p{0.9cm} | >{\centering}p{1.2cm}|>{\centering}p{1.3cm} | >{\centering\arraybackslash}p{1.4cm}|}
\hline
\multicolumn{9}{|c|}{PREDICTIVE ACCURACY FOR TESTING DATASETS 3} \\
\hline
Centrality bin & $\sqrt{s_{NN}}$ [TeV] & Ini. Cond. & $\tau_0$ $(\mathrm{fm/c})$ &$\eta/s$  & $T_{sw}$ [MeV] & L-EOS  & Q-EOS & Accuracy \\
\hline
0\%-50\%   &Au+Au 0.2 & MCKLN   &0.5  & 0.08 & 160   & 128 & 128 & 82\% \\
\hline
0\%-50\%   &Au+Au 0.2 & MC-G   &0.5  & 0.08 & 160   & 128 &128  & 82.28\% \\
\hline
0\%-50\%   &Pb+Pb 2.76  & MC-G   &0.6  & 0.08 & 160  & 256 & 256 & 85\% \\
\hline
0\%-50\%   &Pb+Pb 2.76 & MC-G   &0.4  & 0.16 & 155 & 118 &118 & 84.32\% \\
\hline
\end{tabular}
\caption{Predictive accuracy on testing datasets 3: cascade-coarse-grained spectra $\rho_c(p_T, \Phi)$ generated with the different $\sqrt{s_{NN}}$, initial conditions, $\eta/s$, $\tau_0$, and $T_{sw}$ in the centrality range 0-50\%.} \label{test_sample}
\end{table*}

\begin{table*}[!htbp]
\centering
\begin{tabular}{| >{\centering}p{1.6cm} | >{\centering}p{2.1cm} | >{\centering}p{1.3cm} | >{\centering}p{0.9cm} | >{\centering}p{0.7cm}| >{\centering}p{0.9cm} | >{\centering}p{1.2cm}|>{\centering}p{1.3cm} | >{\centering\arraybackslash}p{1.4cm}|}
\hline
\multicolumn{9}{|c|}{PREDICTIVE ACCURACY FOR TESTING DATASETS 4} \\
\hline
Centrality bin & $\sqrt{s_{NN}}$ [TeV] & Ini. Cond. & $\tau_0$ $(\mathrm{fm/c})$ &$\eta/s$  & $T_{sw}$ [MeV]& L-EOS  & Q-EOS & Accuracy  \\
\hline
15\%-16\%   &Au+Au 0.2  & MCKLN   &0.6 & 0.16    & 160 & 640  & 640 & 95.59\%\\
\hline
10\%-11\%   &Au+Au 0.2  & MC-G   &0.4 & 0.12    & 160 &  2560 & 2560 & 100\%\\
\hline
15\%-16\%   &Au+Au 0.2  & MC-G   &0.4 & 0.12    & 160 &  2560 & 2560  & 99.8\% \\
\hline
20\%-21\%   &Au+Au 0.2  & MC-G   &0.4 & 0.12    & 160 &  2560& 2560 & 94.9\%\\
\hline
15\%-16\%   &Au+Au 0.2  & MC-G   &0.4 & 0.00    & 155 &  2560& 2560& 74.86\%\\
\hline
20\%-21\%   &Au+Au 0.2  & MC-G   &0.4 & 0.12    & 155 &  1792 & 1792 & 88.8\% \\
\hline
15\%-16\%   &Pb+Pb 2.76 & MC-G  &0.6  & 0.08    & 155 &  2560& 2560 & 99.99\%\\
\hline
20\%-21\%   &Pb+Pb 2.76 & MC-G  &0.6  & 0.08    & 155 &  2560& 2560 & 99.78\%\\
\hline
\end{tabular}
\caption{Predictive accuracy on testing datasets 4: 30-events-fine-averaged spectra $\rho_a(p_T, \Phi)$ generated with the different $\sqrt{s_{NN}}$, initial conditions, $\eta/s$, $\tau_0$, and $T_{sw}$ in the centrality range 0-30\%.} \label{test-data-central-155}
\end{table*}

\begin{table*}[!htbp]
\centering
\begin{tabular}{| >{\centering}p{1.6cm} | >{\centering}p{2.1cm} | >{\centering}p{1.3cm} | >{\centering}p{0.9cm} | >{\centering}p{0.9cm}| >{\centering}p{0.9cm} | >{\centering}p{1.2cm}|>{\centering}p{1.3cm} | >{\centering\arraybackslash}p{1.4cm}|}
\hline
\multicolumn{9}{|c|}{PREDICTIVE ACCURACY FOR TESTING DATASETS} \\
\hline
Centrality bin & $\sqrt{s_{NN}}$ [TeV] & Ini. Cond. & $\tau_0$ $(\mathrm{fm/c})$ &$\eta/s(T)$  & $T_{sw}$ [MeV]& L-EOS  & Q-EOS & Accuracy  \\
\hline
10\%-11\%   &Au+Au 0.2  & MC-G   &0.4 & 1         & 160 & 512  & 512 & 100\%\\
\hline
10\%-11\%   &Au+Au 0.2  & MC-G   &0.4 & 2         & 160 & 512  & 512 & 100\%\\
\hline
10\%-11\%   &Au+Au 0.2  & MC-G   &0.4 & 3         & 160 & 512  & 512 & 100\%\\
\hline
10\%-11\%   &Au+Au 0.2  & MC-G   &0.4 & 4         & 160 & 512  & 512 & 100\%\\
\hline
15\%-16\%   &Au+Au 0.2  & MC-G   &0.4 & 4        & 160 &  512 & 512 &   99.51\% \\
\hline
20\%-21\%   &Au+Au 0.2  & MC-G   &0.4 & 3         & 160 &  512 & 512&   98.34\%\\
\hline
10\%-11\%   &Au+Au 0.2  & MCKLN  &0.6 & 3       & 160 &  512 & 512 & 98.04\% \\
\hline
10\%-11\%   &Pb+Pb 2.76 & MC-G  &0.6  & 3       & 155 &  512& 512 &  99.80\%\\
\hline
25\%-26\%   &Pb+Pb 2.76 & MC-G  &0.6  & 4       & 160 &  512& 512 &  99.90\%\\
\hline
35\%-36\%   &Pb+Pb 2.76 & MC-G  &0.6  & 3       & 155 &  512& 512 &  86.72\%\\
\hline
\end{tabular}
\caption{Predictive accuracy on testing datasets 5: 30-events-fine-averaged spectra $\rho_a(p_T, \Phi)$ generated with the different $\sqrt{s_{NN}}$, initial conditions,  temperature-dependent $\eta/s(T)$, $\tau_0$, and $T_{sw}$ in the centrality range 0-30\%.} \label{test-data-central-155-viscosity}
\end{table*}

\begin{figure}[bt]
\centering
\includegraphics[width=0.50\textwidth]{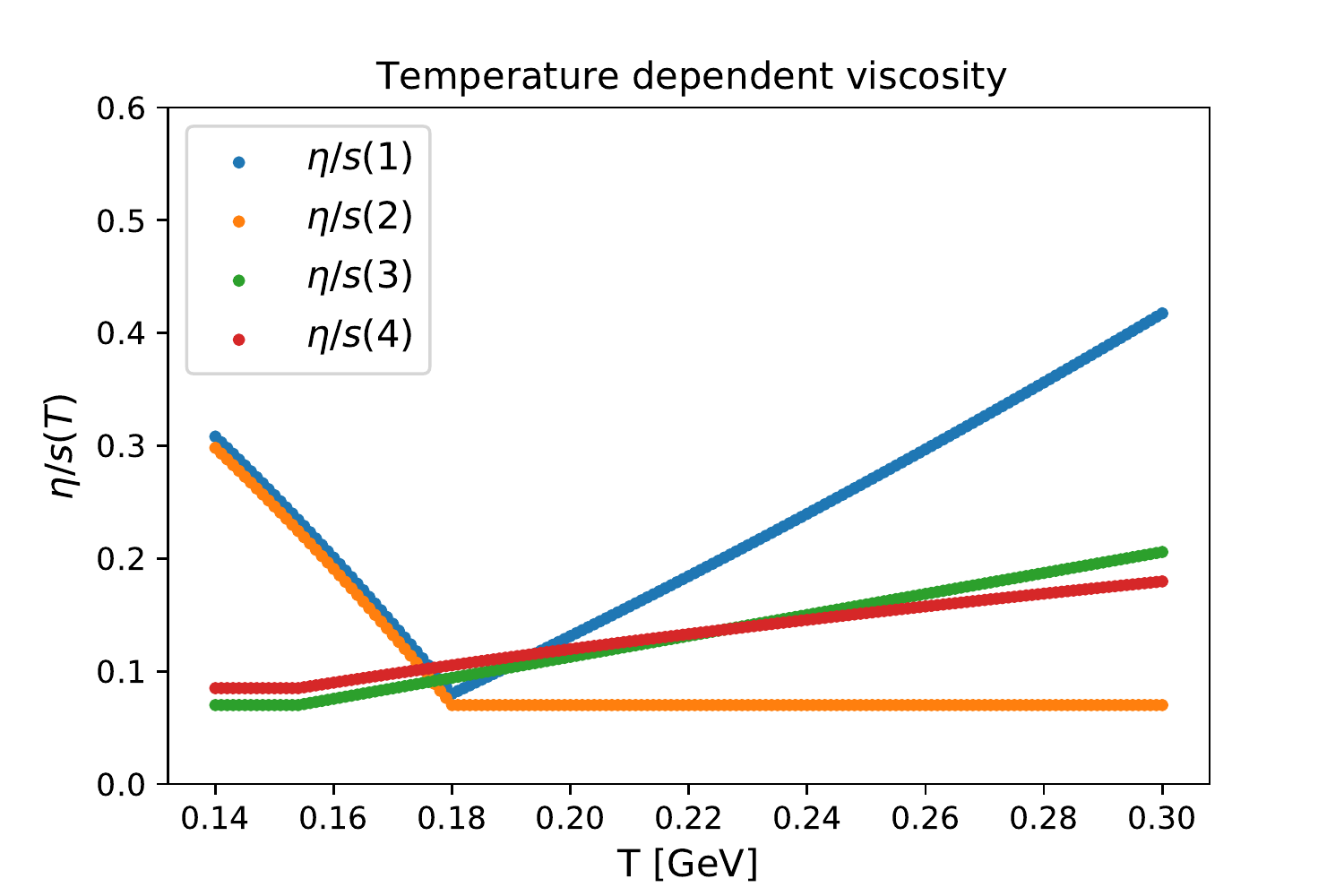} 
\caption{Four temperature-dependent viscosities from Ref.~\cite{niemi2011influence,bernhard2019bayesian,bernhard2016applying}. The $\eta/s(2)$ orange line is shifted downwards for better visibility.}
\label{Fig: Temperature_dependent_viscosity}
\end{figure}

\section{Traditional observables from the training data}
\label{appendix: traditional observables}
\begin{figure*}[thbp]
\centering
\includegraphics[width=0.48\textwidth]{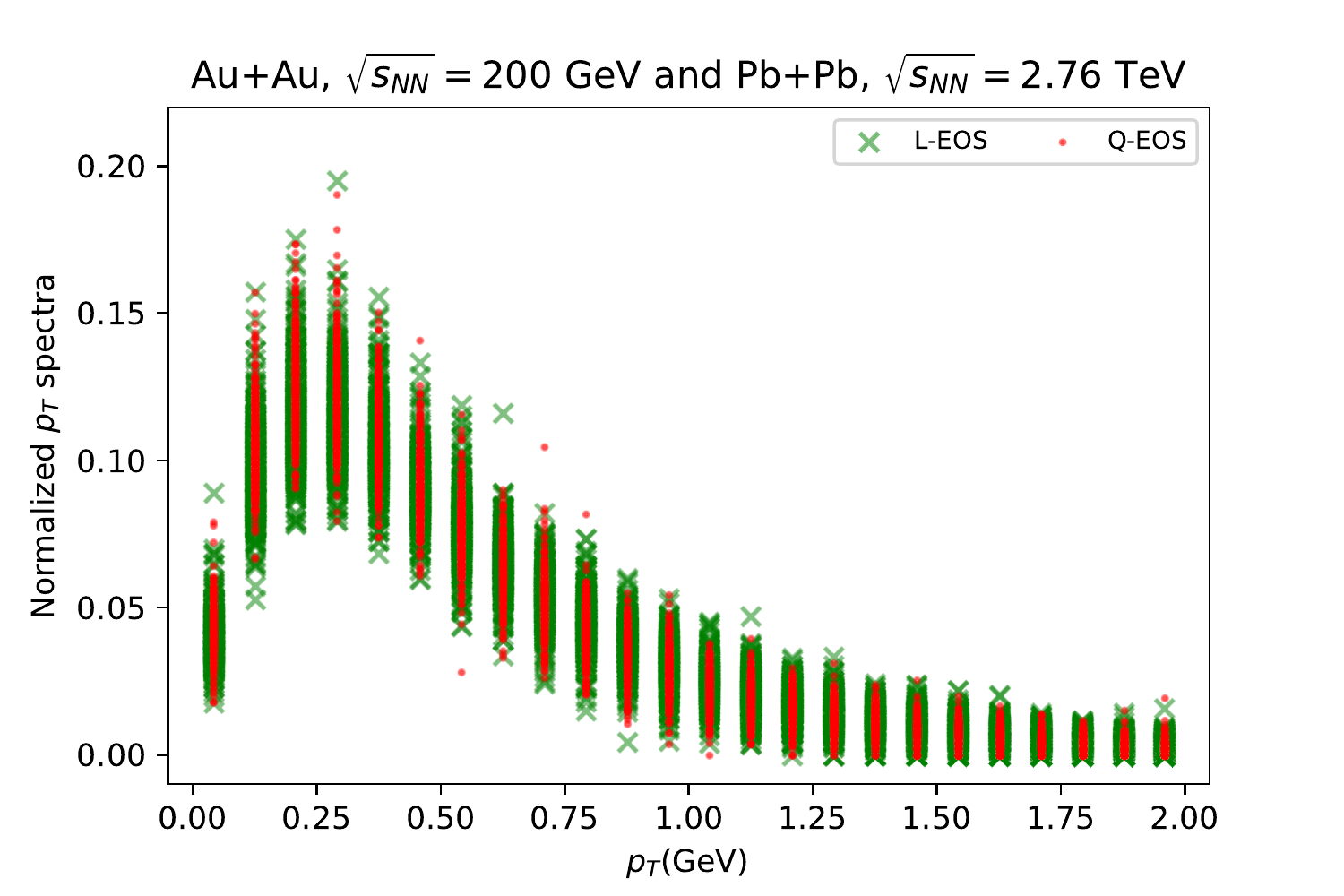} 
\includegraphics[width=0.48\textwidth]{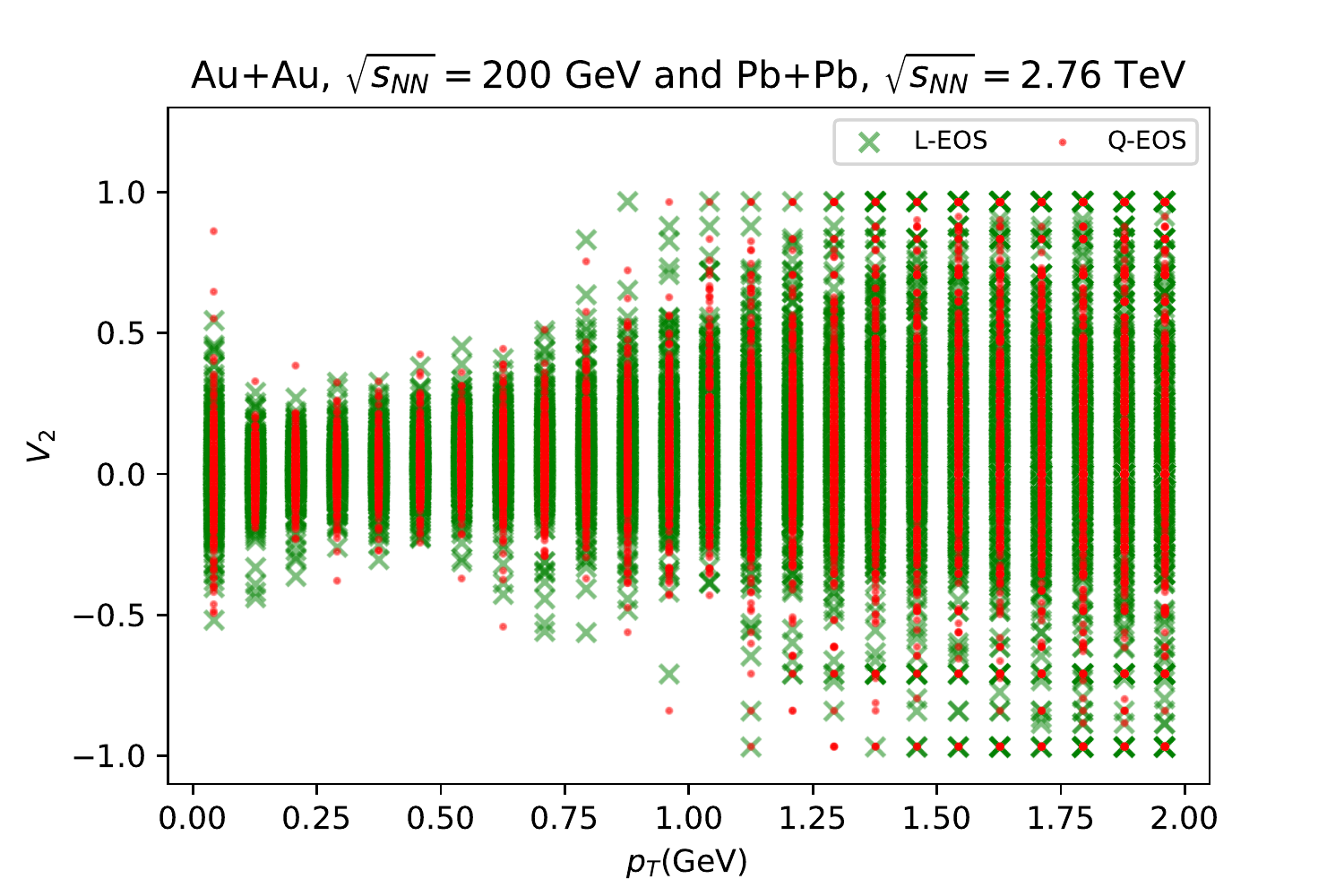} 
\caption{Event-by-event normalized $p_T$ spectra  $\mathrm{d}N/N\mathrm{d}y\mathrm{d}p_T$ (left panel) and elliptic flow $v_2$ as a function of $p_T$ (right panel) of the training datasets in Tab.~\ref{data 2760} and Tab.~\ref{data 200} with two EoSs. Vertical discrepancy is event-by-event fluctuations. The green cross and the red point symbol depict the observables with L-EOS and Q-EOS, respectively. These events are generated in different centrality bins with $T_{sw}=137$ MeV in two collision systems.}
\label{fig-PTV2_spectra_ebe}
\end{figure*}

\begin{figure*}[htbp]
\centering
\includegraphics[width=0.48\textwidth]{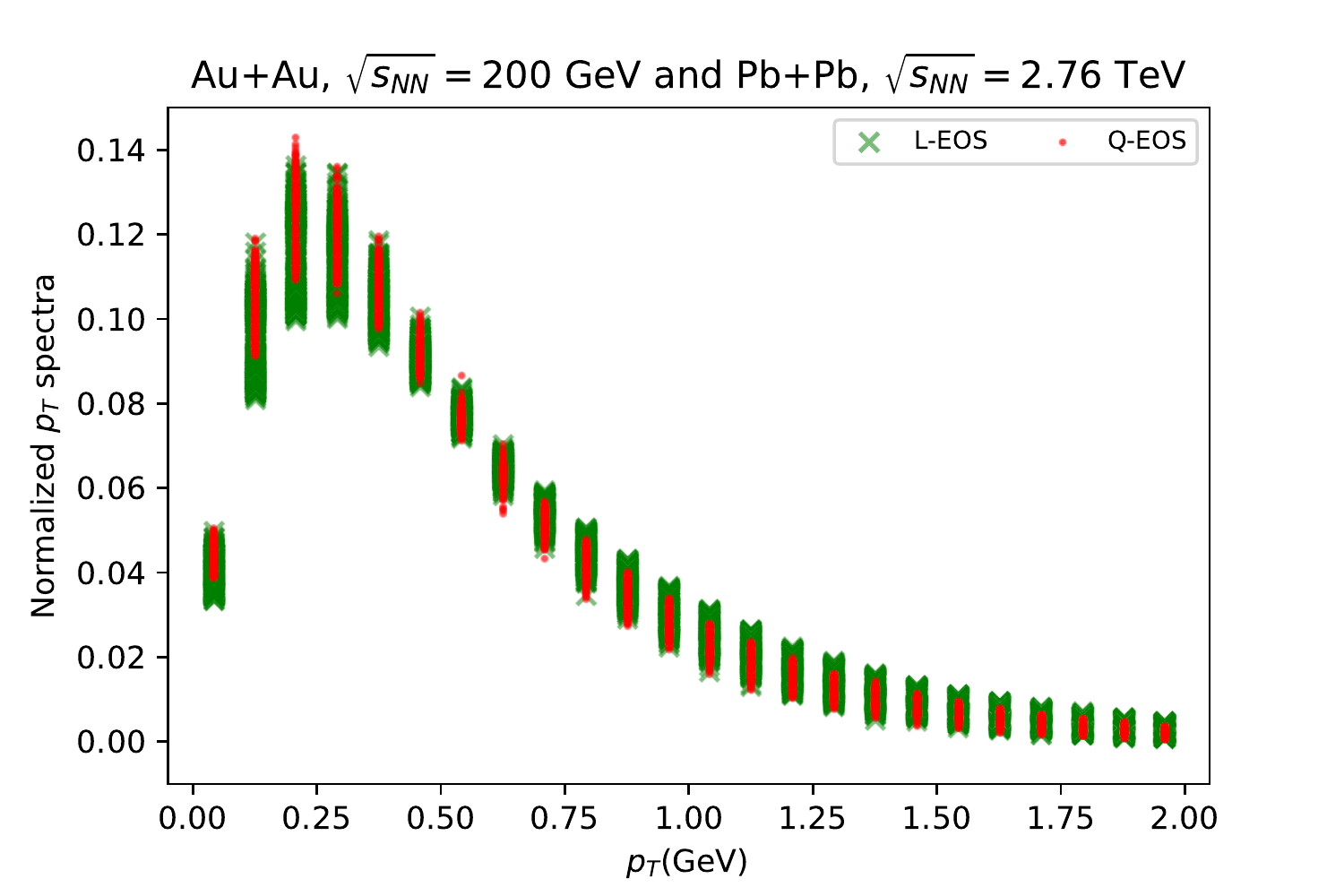} 
\includegraphics[width=0.48\textwidth]{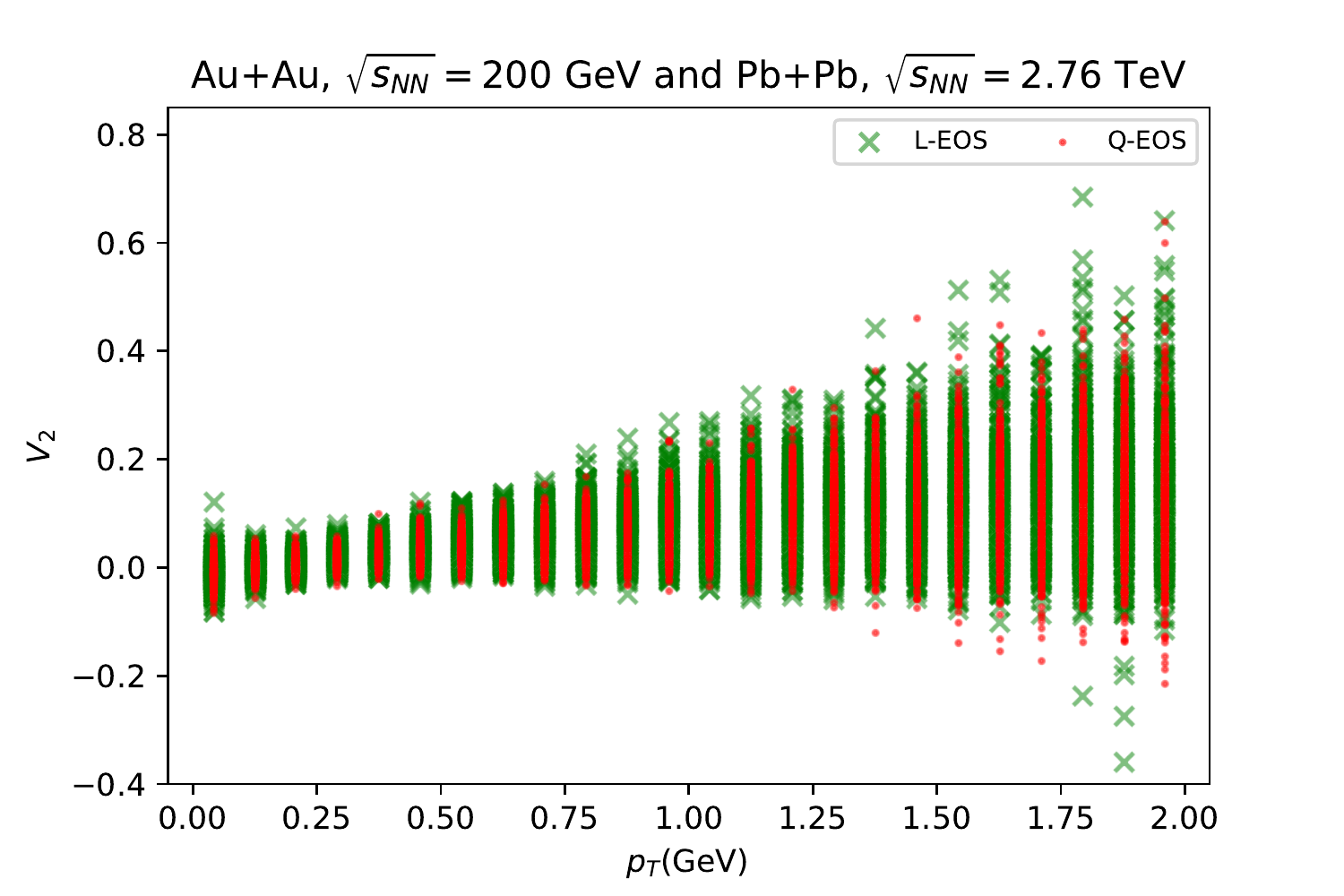} 
\caption{Same as Fig.~\ref{fig-PTV2_spectra_ebe} but for 30-events-fine-averaged normalized $p_T$ spectra $\mathrm{d}N/N\mathrm{d}y\mathrm{d}p_T$ (left panel) and elliptic flow $v_2$ as a function of $p_T$ (right panel).}
\label{fig-PTV2_spectra_fine}
\end{figure*}

\begin{figure*}[thbp]
\centering
\includegraphics[width=0.48\textwidth]{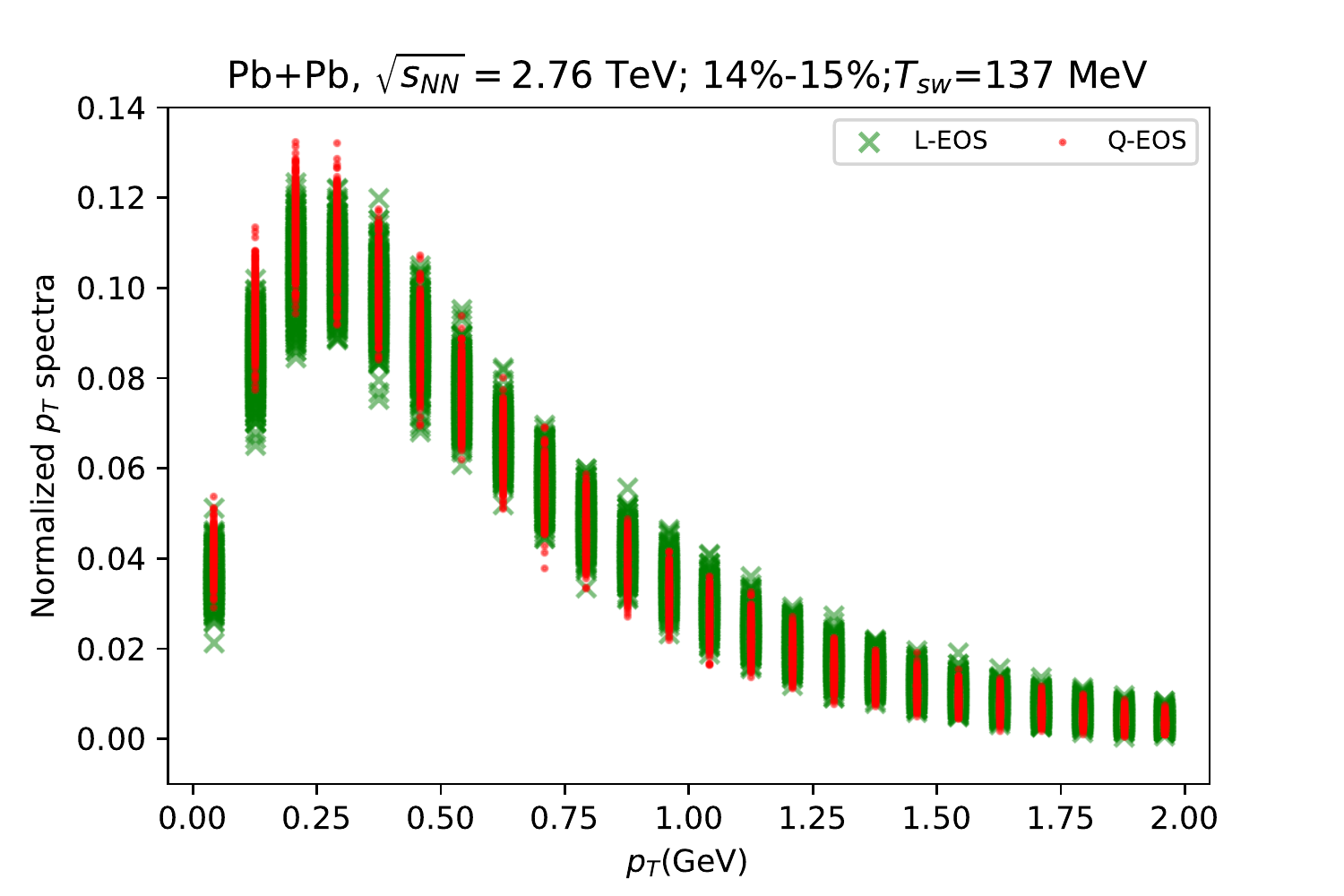} 
\includegraphics[width=0.48\textwidth]{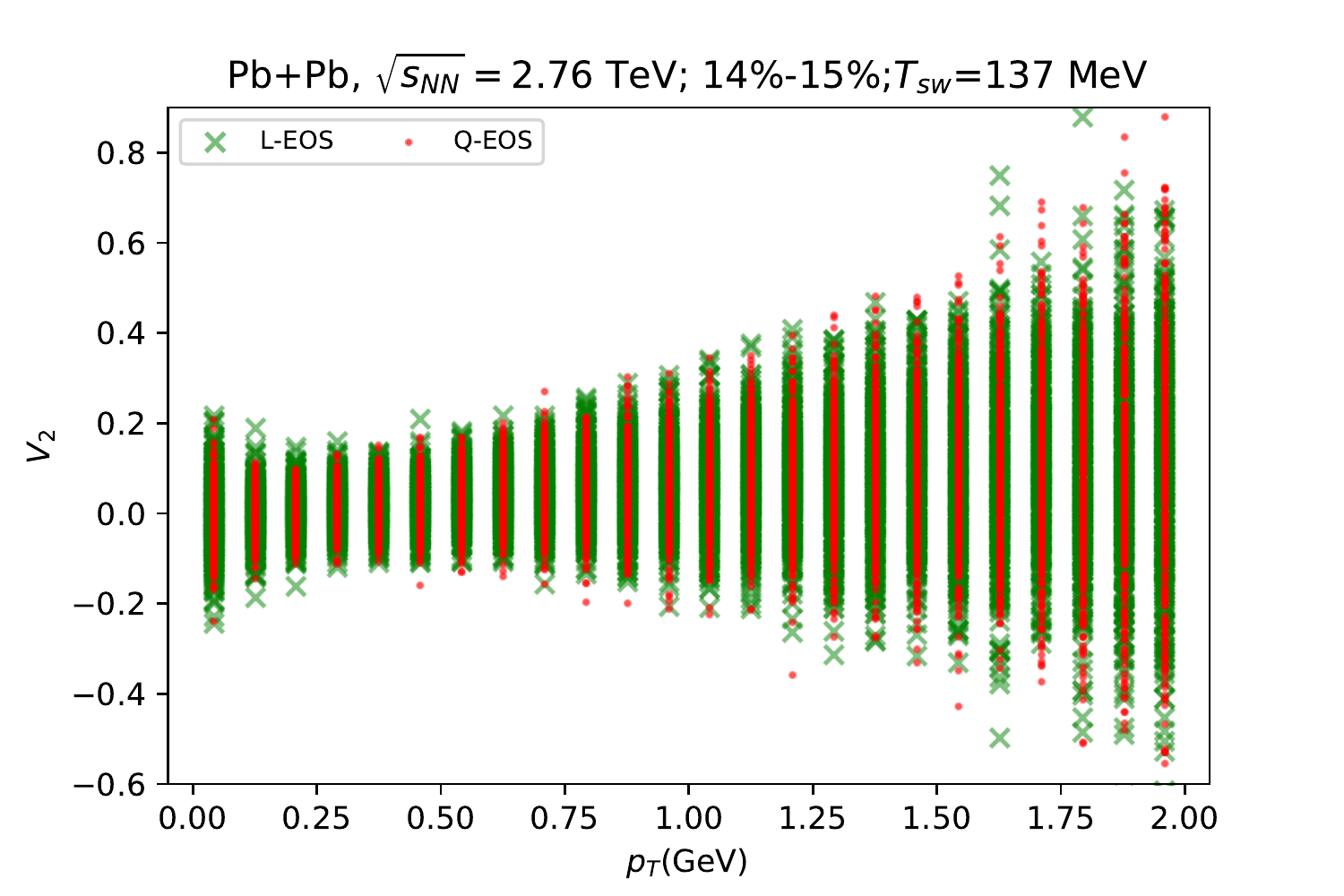} 
\caption{Event-by-event normalized $p_T$ spectra  $\mathrm{d}N/N\mathrm{d}y\mathrm{d}p_T$ (left panel) and elliptic flow $v_2$ as a function of $p_T$ (right panel) of the training datasets in Tab.~\ref{data 2760} with two EoSs. The green cross and the red point symbol depict the observables with L-EOS and Q-EOS, respectively. These events are generated in centrality bin 14\%-15\% with $T_{sw}=137$ MeV in two collision systems.}
\label{fig-PTV2_spectra_ebe_onebin}
\end{figure*}

\begin{figure*}[htbp]
\centering
\includegraphics[width=0.48\textwidth]{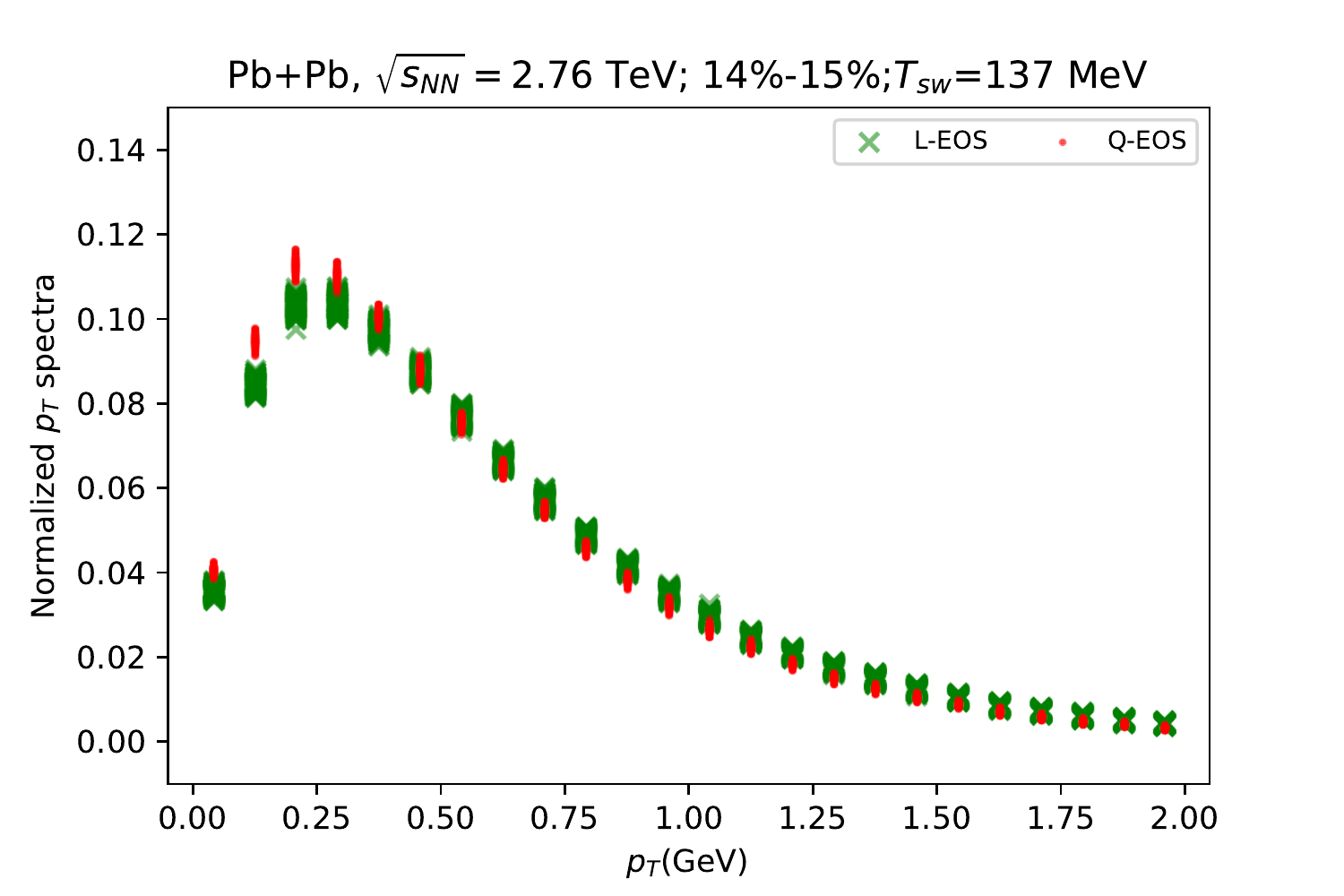} \includegraphics[width=0.48\textwidth]{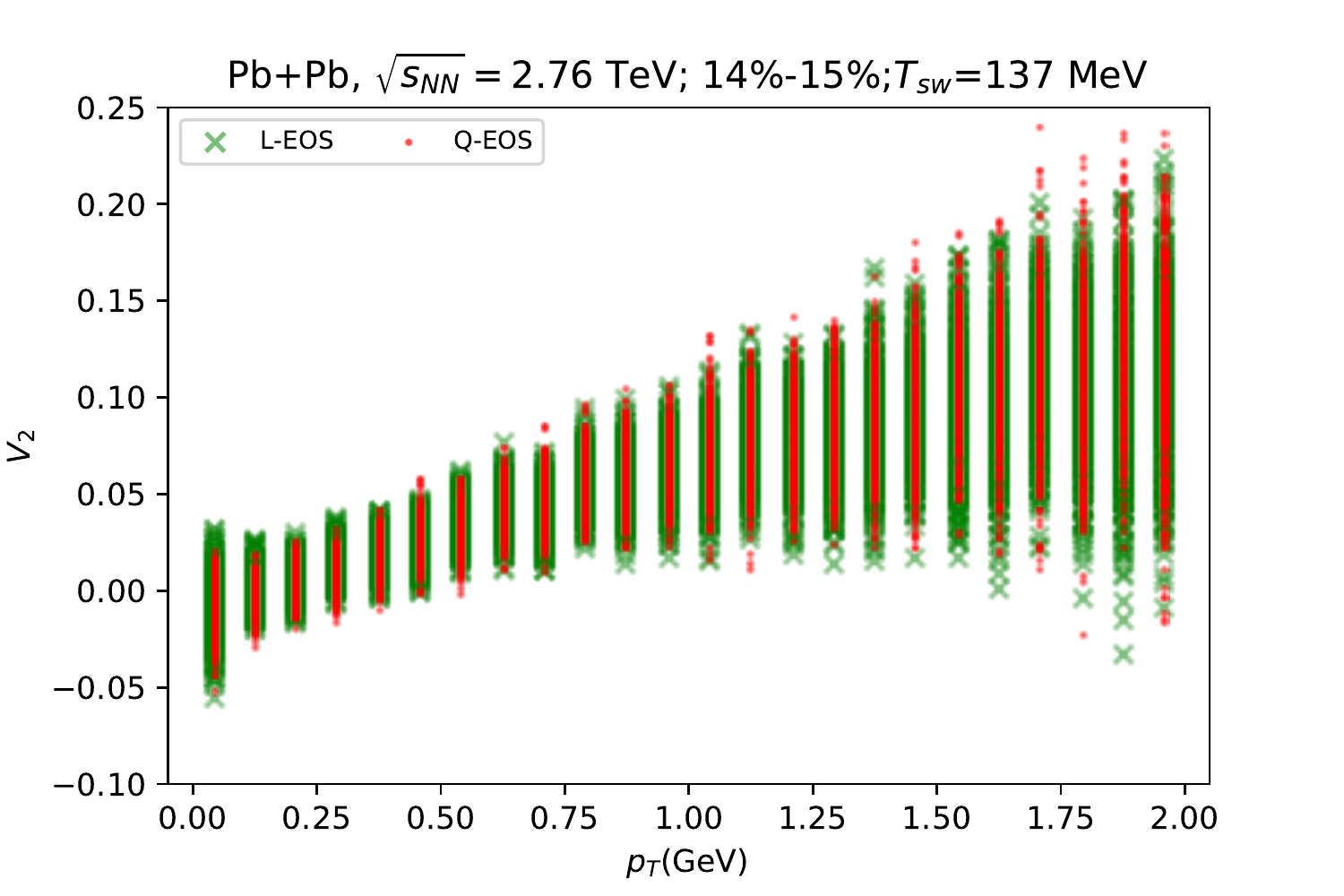} \caption{Same as Fig.~\ref{fig-PTV2_spectra_ebe_onebin} but for 30-events-fine-averaged normalized $p_T$ spectra $\mathrm{d}N/N\mathrm{d}y\mathrm{d}p_T$ (left panel) and elliptic flow $v_2$ as a function of $p_T$ (right panel).}
\label{fig-PTV2_spectra_fine_onebin}
\end{figure*}

\begin{figure*}[thbp]
\centering
\includegraphics[width=0.48\textwidth]{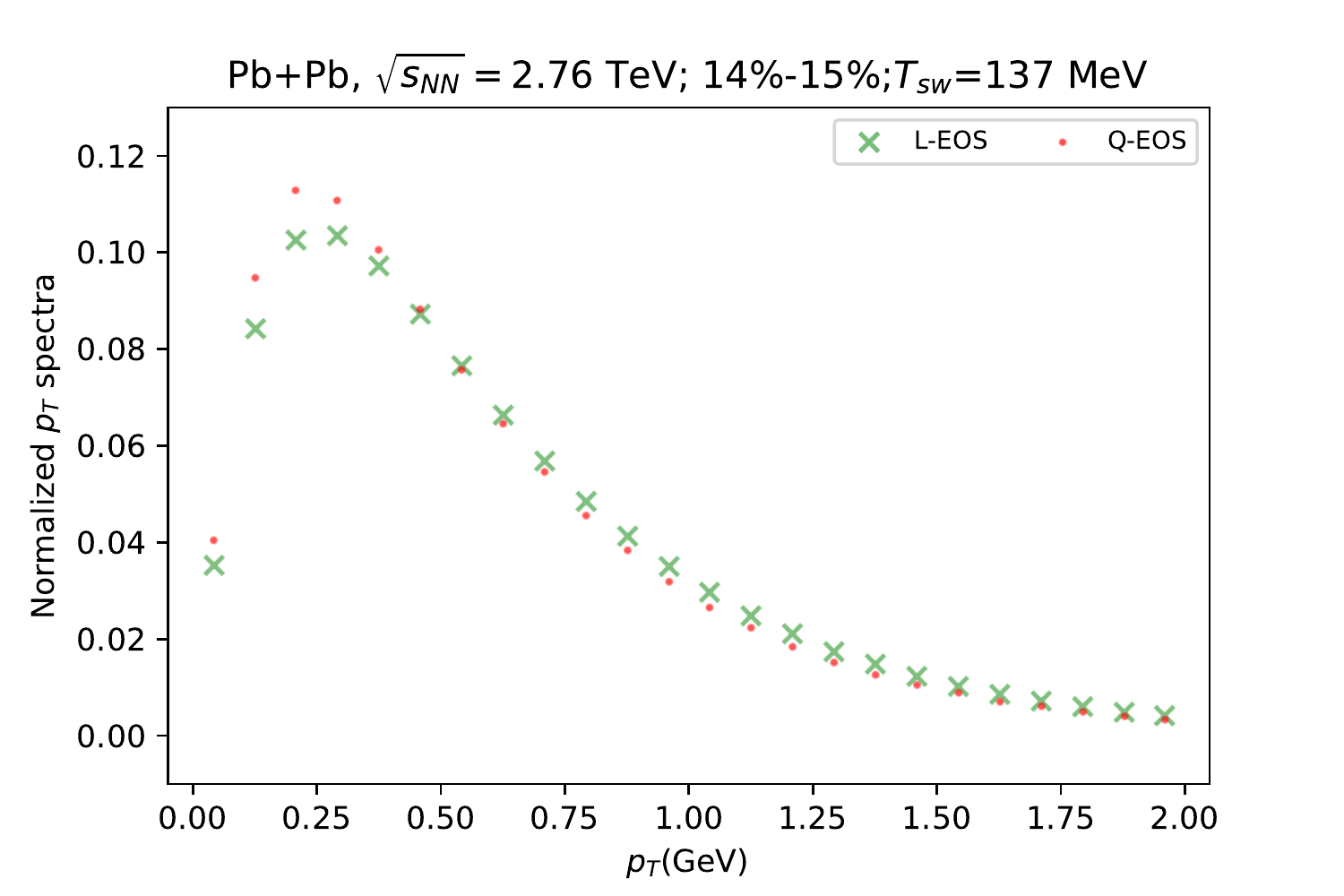} 
\includegraphics[width=0.48\textwidth]{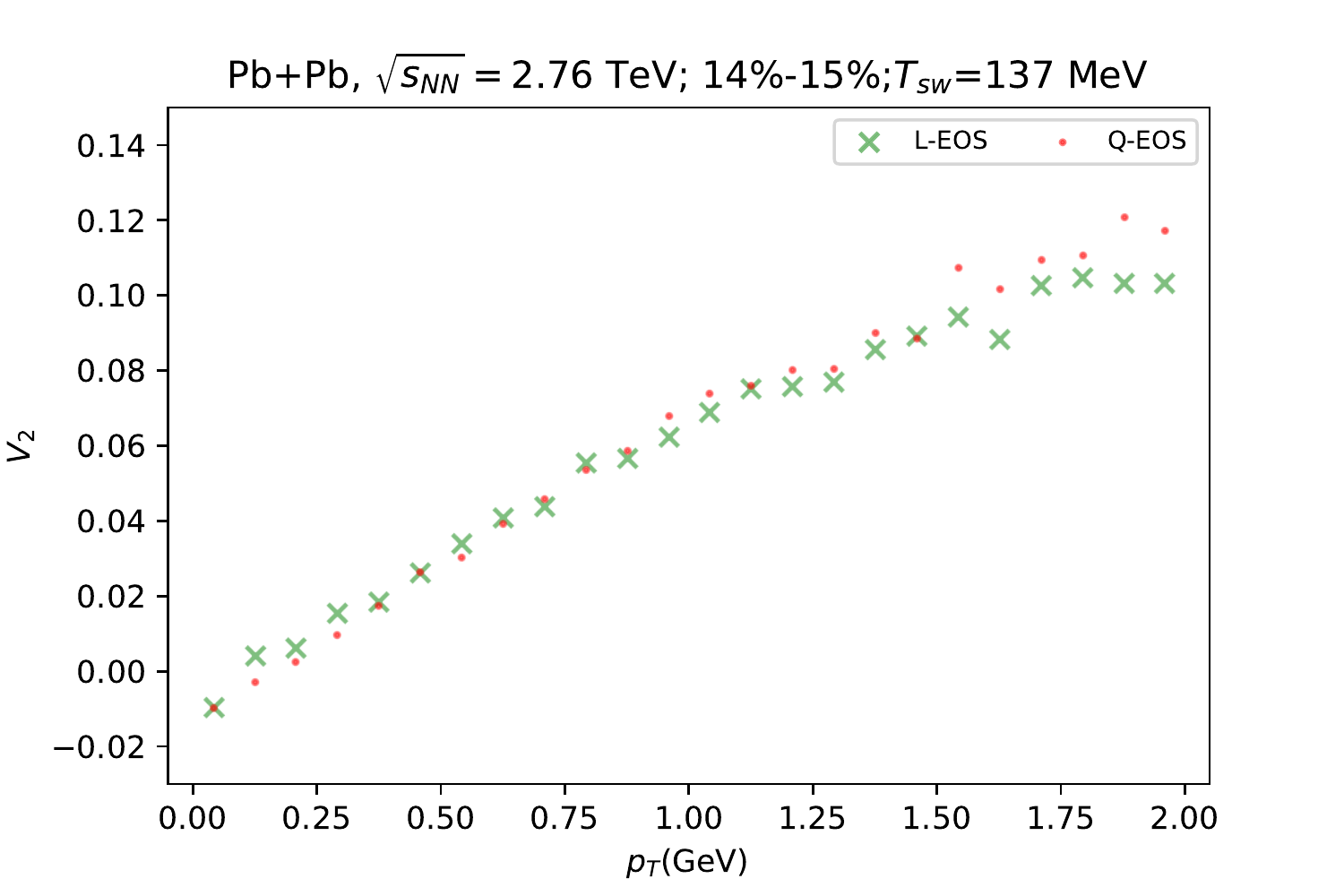} 
\caption{All-events-fine-averaged normalized $p_T$ spectra  $\mathrm{d}N/N\mathrm{d}y\mathrm{d}p_T$ (left panel) and elliptic flow $v_2$ as a function of $p_T$ (right panel) of the training datasets in Tab.~\ref{data 2760} with two EoSs. The green cross and the red point symbol depict the observables with L-EOS and Q-EOS, respectively. These events are generated in centrality bin 14\%-15\% with $T_{sw}=137$ MeV in two collision systems.}
\label{fig-PTV2_spectra_mean_onebin}
\end{figure*}

\begin{figure*}[thbp]
\centering
\includegraphics[width=0.48\textwidth]{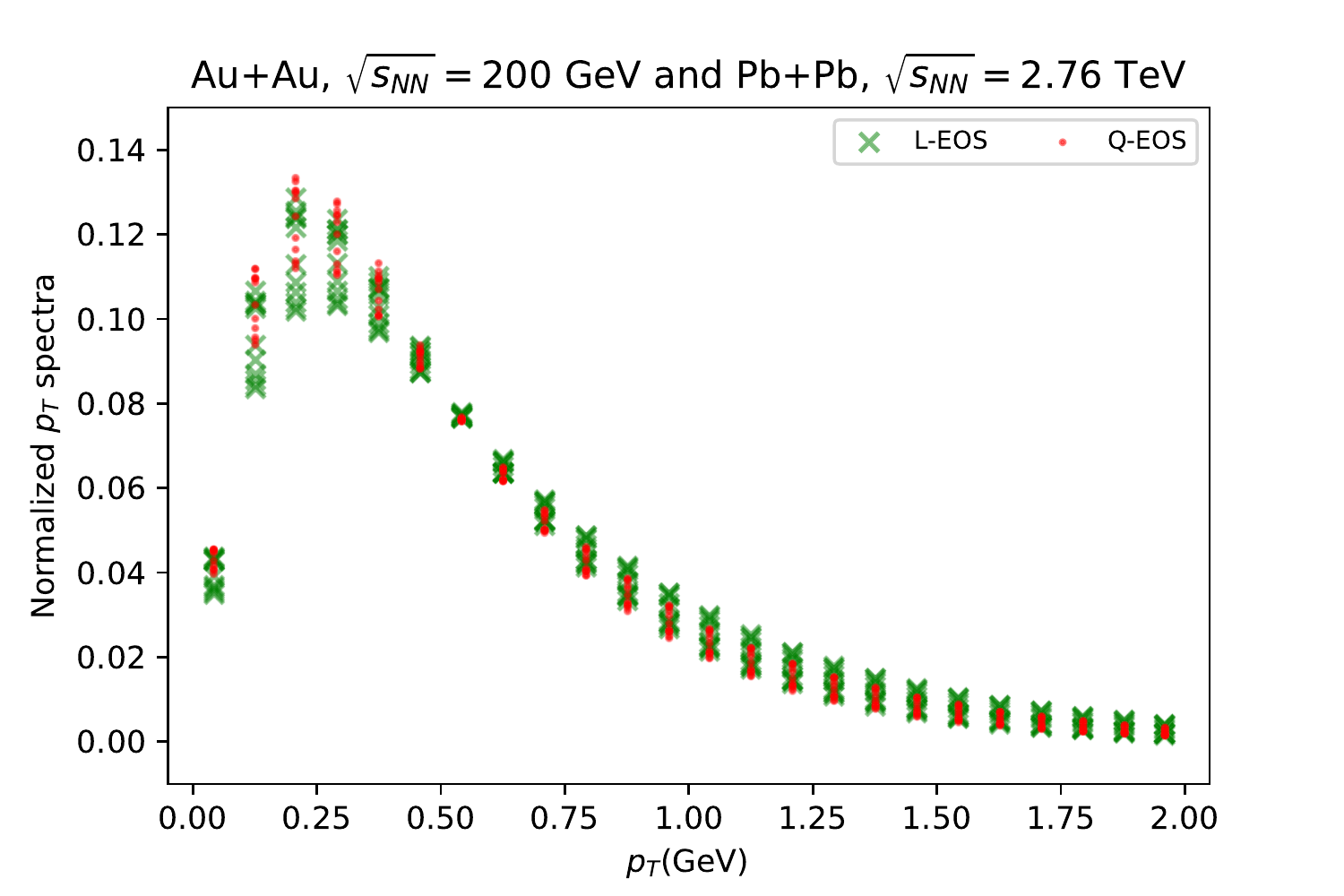} 
\includegraphics[width=0.48\textwidth]{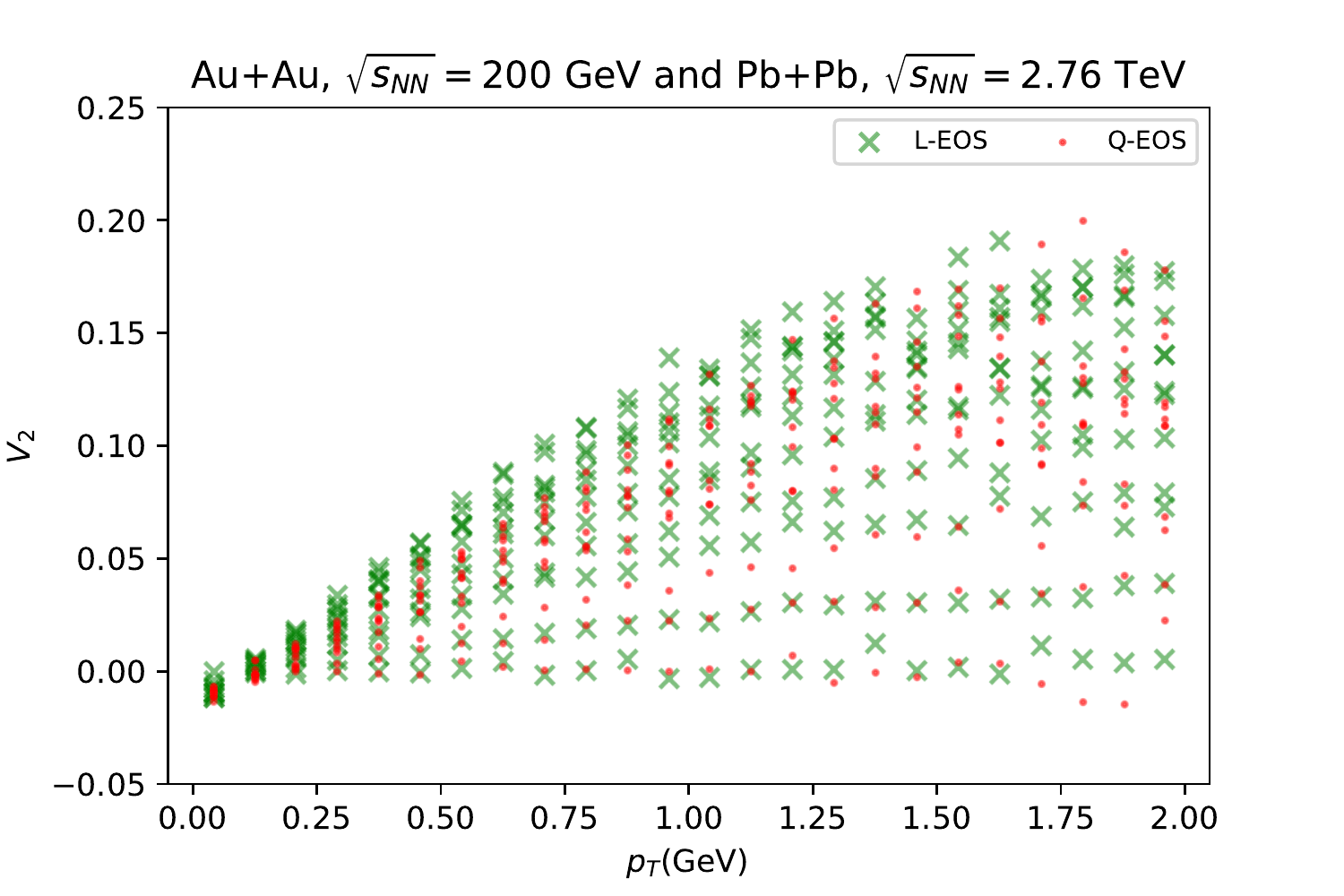}\\
\includegraphics[width=0.32\textwidth]{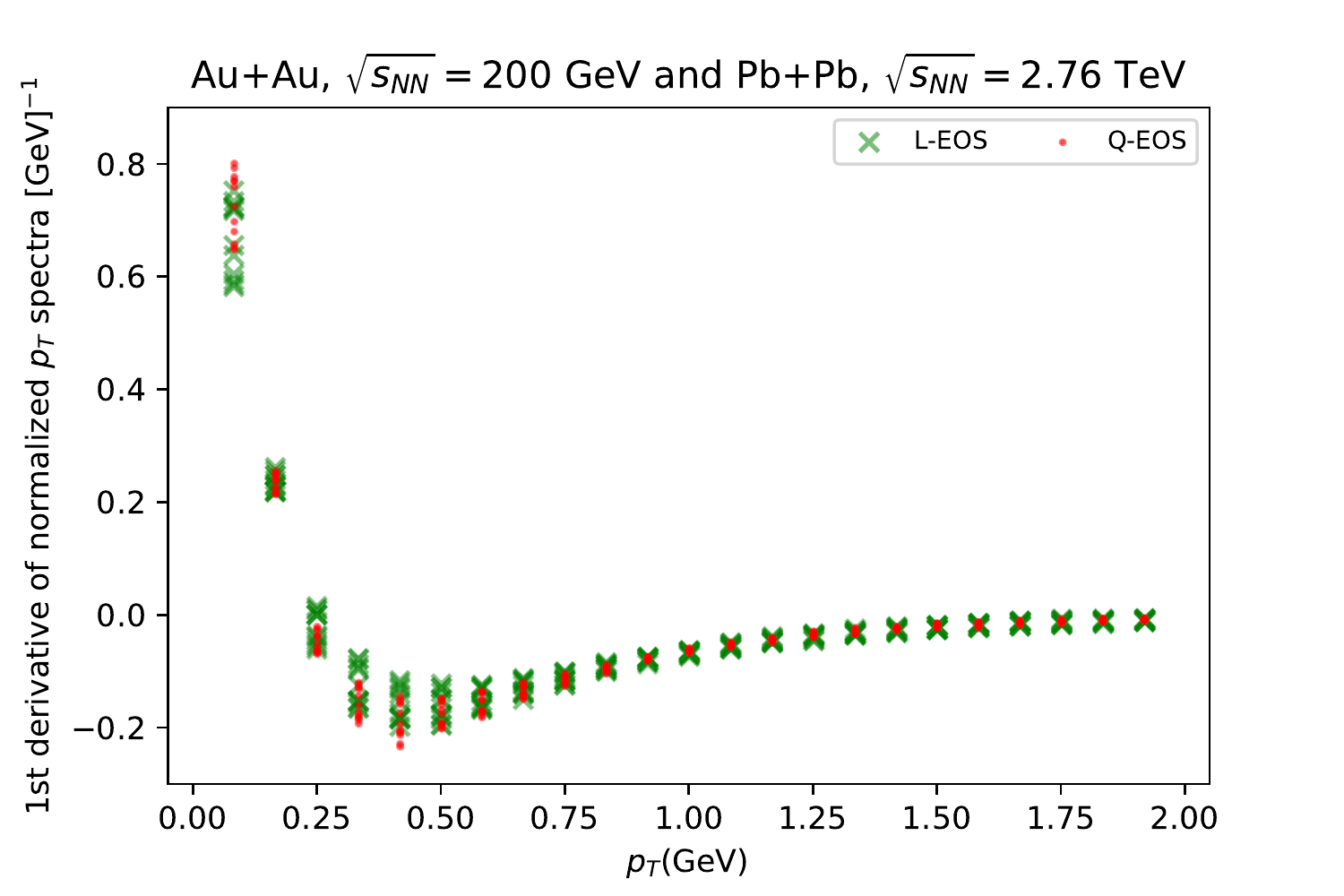} 
\includegraphics[width=0.32\textwidth]{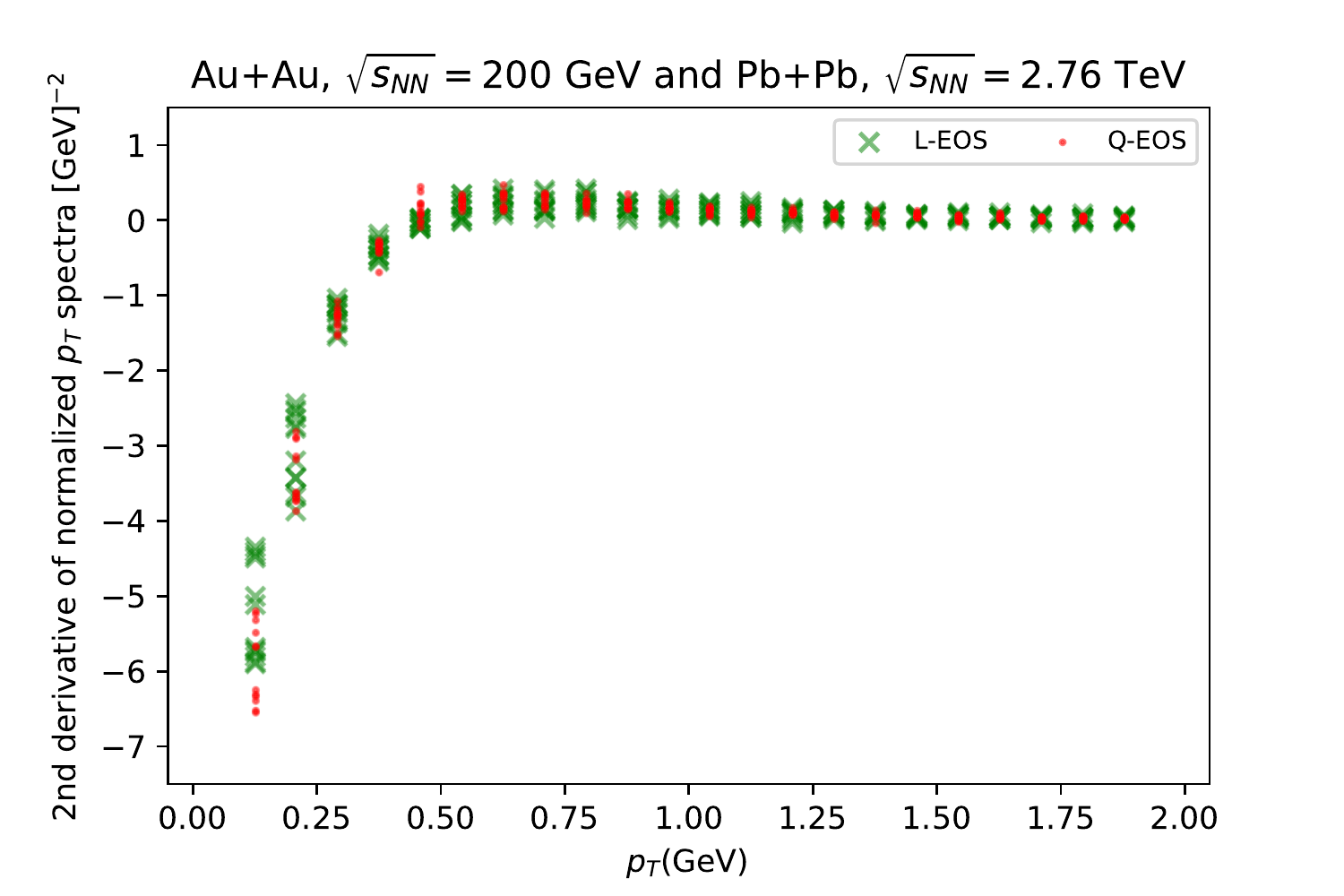} 
\includegraphics[width=0.32\textwidth]{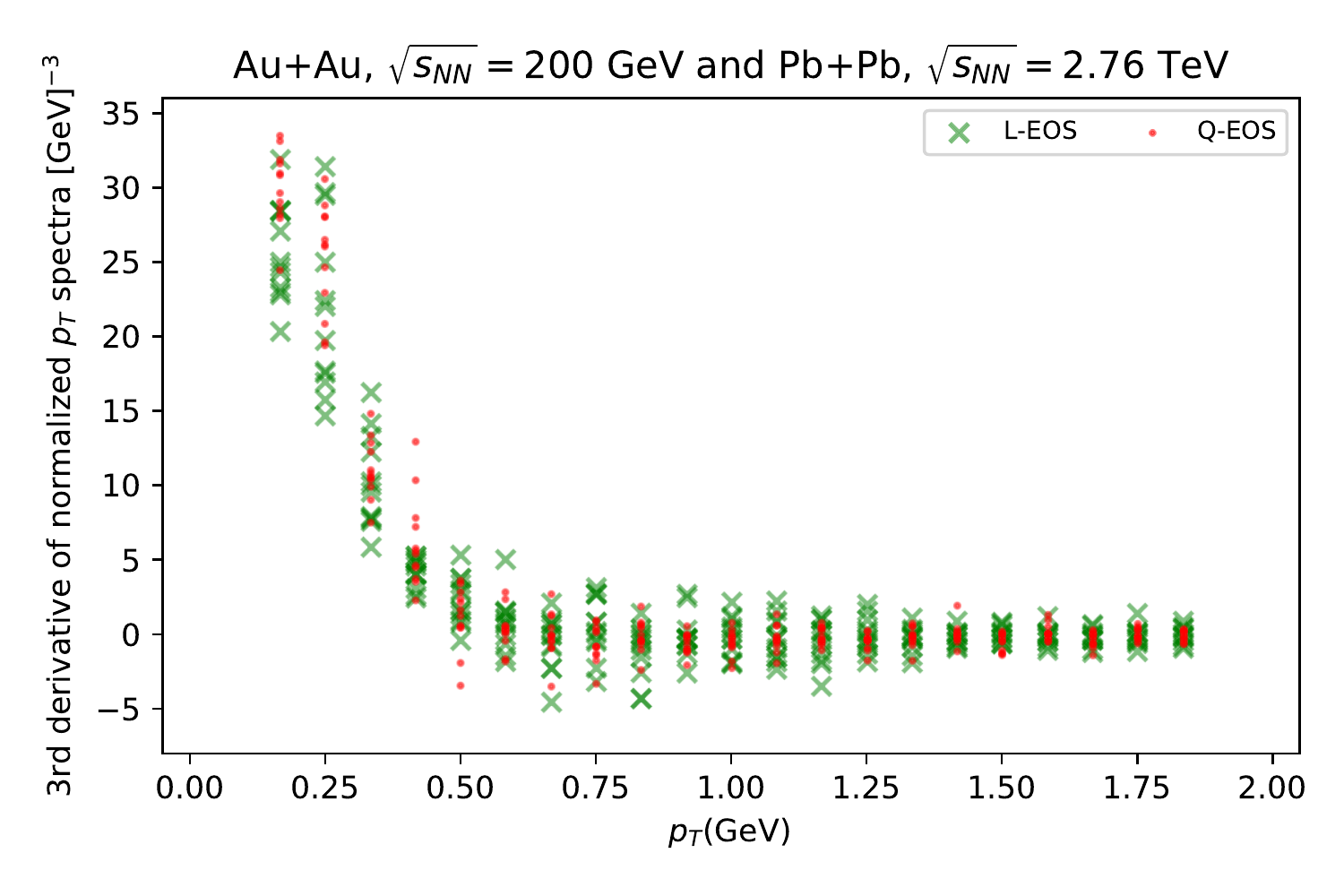} 

\caption{All-events-fine-averaged normalized $p_T$ spectra $\mathrm{d}N/N\mathrm{d}y\mathrm{d}p_T$ (upper left panel) and elliptic flow $v_2$ as a function of $p_T$ (upper right panel) and the first, second and third derivative of these normalized $p_T$ spectra (lower panel) of the training datasets in Tab.~\ref{data 2760} and Tab.~\ref{data 200} with two EoSs. The green cross and the red point symbol depict the observables with L-EOS and Q-EOS, respectively. These events are generated in different centrality bins with $T_{sw}=137$ MeV in two collision systems.}
\label{fig-PTV2_spectra_mean}
\end{figure*}

\section*{Bibliography}
\bibliographystyle{apsrev4-1}
\bibliography{duyl}
\end{document}